\documentclass[preprint2]{aastex}

\slugcomment{Comment}

\shorttitle{Reborn X-ray Emission from the Born-Again PN A\,30}
\shortauthors{Guerrero et al.}

\begin{document}


\title{Rebirth of X-ray Emission from the Born-Again Planetary Nebula A\,30} 


\author{M.A.\ Guerrero\altaffilmark{1}, 
N.\ Ruiz\altaffilmark{1}, 
W.-R.\ Hamann\altaffilmark{2}, 
Y.-H.\ Chu\altaffilmark{3}, 
H.\ Todt\altaffilmark{2}, 
D.\ Sch\"onberner\altaffilmark{4}, 
L.\ Oskinova\altaffilmark{2}, 
R.A.\ Gruendl\altaffilmark{3}, 
M.\ Steffen\altaffilmark{4}, 
W.P.\ Blair\altaffilmark{5}, 
\& 
J.A.\ Toal\'a\altaffilmark{1}
}

\email{mar@iaa.es}


\altaffiltext{1}{Instituto de Astrof\'\i sica de Andaluc\'\i a, IAA-CSIC, 
c/ Glorieta de la Astronom\'\i a s/n, 18008 Granada, Spain}
\altaffiltext{2}{Institute for Physics and Astronomy, Universit\"at Potsdam, 
14476 Potsdam, Germany}
\altaffiltext{3}{Department of Astronomy, University of Illinois, 
1002 West Green Street, Urbana, IL 61801, USA} 
\altaffiltext{4}{Leibniz-Institut f\"ur Astrophysik Potsdam (AIP), 
An der Sternwarte 16, 14482 Potsdam, Germany} 
\altaffiltext{5}{Department of Physics and Astronomy, Johns Hopkins University, 
Baltimore, MD 21218, USA}


\begin{abstract}

The planetary nebula (PN) A\,30 is believed to have undergone a very late 
thermal pulse resulting in the ejection of knots of hydrogen-poor material.  
Using multi-epoch \emph{HST} images we have detected the angular 
expansion of these knots and derived an age of 850$^{+280}_{-150}$ 
yr.  
To investigate the spectral and spatial properties of the soft X-ray 
emission detected by \emph{ROSAT}, we have obtained \emph{Chandra} 
and \emph{XMM-Newton} deep observations of A\,30.  
The X-ray emission from A\,30 can be separated into two components: 
a point-source at the central star and diffuse emission associated 
with the hydrogen-poor knots and the cloverleaf structure inside the 
nebular shell.  
To help us assess the role of the current stellar wind in powering this 
X-ray emission, we have determined the stellar parameters and wind 
properties of the central star of A\,30 using a non-LTE model fit to its 
optical and UV spectrum.  
The spatial distribution and spectral properties of the diffuse X-ray 
emission is highly suggestive that it is generated by the post-born-again 
and present fast stellar winds interacting with the hydrogen-poor ejecta 
of the born-again event.  
This emission can be attributed to shock-heated plasma, as the 
hydrogen-poor knots are ablated by the stellar winds, under which 
circumstances the efficient mass-loading of the present fast 
stellar wind raises its density and damps its velocity to produce 
the observed diffuse soft X-rays.
Charge transfer reactions between the ions of the stellar winds and 
material of the born-again ejecta has also been considered as a 
possible mechanism for the production of diffuse X-ray emission, and 
upper limits on the expected X-ray production by this mechanism have 
been derived.  
The origin of the X-ray emission from the central star of A\,30 is 
puzzling: shocks in the present fast stellar wind and photospheric 
emission can be ruled out, while the development of a new, compact 
hot bubble confining the fast stellar wind seems implausible.  

\end{abstract}





\section{Introduction}

Planetary nebulae (PNe) consist of stellar material ejected by low- and 
intermediate-mass stars. 
In the canonical model of PN formation, the so-called interacting stellar 
winds (ISW) model, the envelope of a star is stripped off through a slow 
and dense wind and, as the star evolves off the asymptotic giant branch 
(AGB), it is subsequently swept up by a fast 
stellar wind \citep{C-SP85} to form a PN \citep{KPF78,B87}.  
The dynamical structure of a PN can be very complex as the fast 
stellar wind interacts with a slow AGB wind whose density and 
velocity structure has been previously modified by the passage 
of the shock wave associated with a D-type ionization front 
\citep[][]{Perinotto_etal04,Schonberner_etal05a,Schonberner_etal05b}.

\begin{figure*}
\includegraphics[width=6.5in]{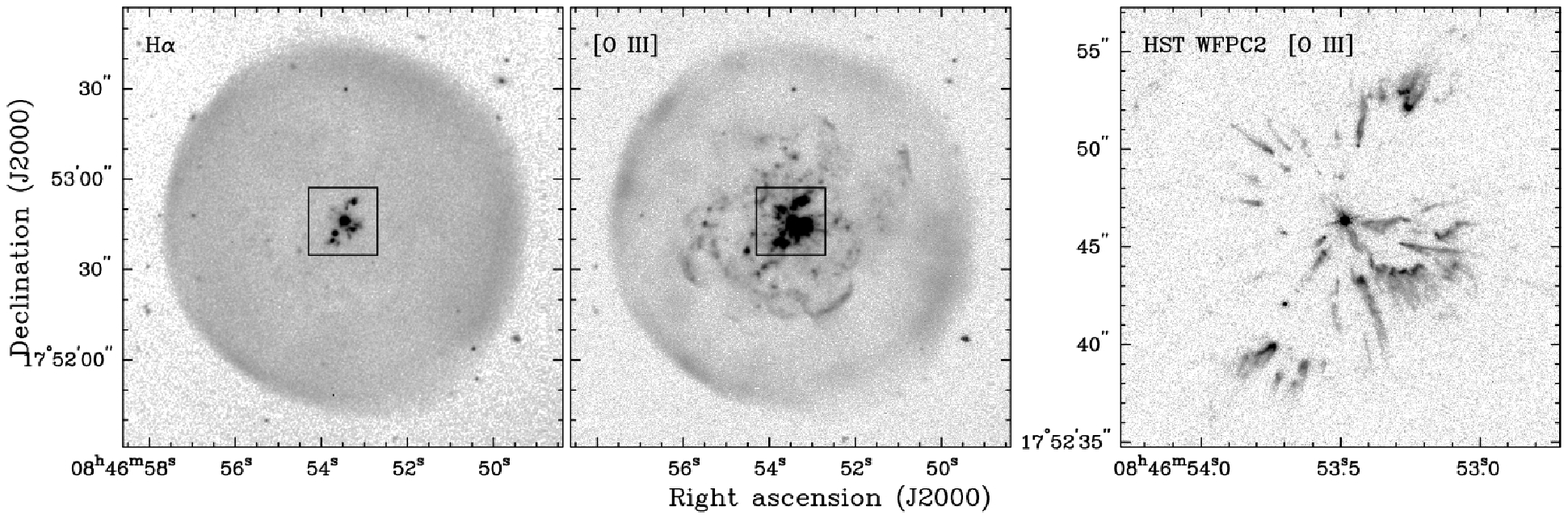}
\caption{
KPNO Mayall 4m CCD H$\alpha$ ({\it left}), [O~{\sc iii}] ({\it center}), 
and \emph{HST} WFPC2 [O~{\sc iii}] ({\it right}) images of A\,30.  
The boxes overlaid in the ground-based images correspond to the field 
of view of the \emph{HST} image shown.  
The ground-based images were acquired through filters with central 
wavelengths 5025 \AA\ and 6580 \AA, and FWHMs 50 \AA\ and 60 \AA, 
respectively;  the \emph{HST} WFPC2 image used the F502N filter.  
}
\label{img.opt}
\end{figure*}

Abell\,30 (a.k.a.\ A\,30, PN\,G208.5+33.2) is a PN with a hydrogen-deficient 
central star (CSPN) of spectral type [WC]-PG1159 (also termed as ``weak 
emission line stars'').  
The nebula appears in H$\alpha$ (Figure~\ref{img.opt}-{\it left}) 
as a limb-brightened, presumably spherical shell with a diameter 
$\sim$2\arcmin, although a close inspection of its kinematics reveals 
that this shell has a mildly ellipsoidal shape \citep{ML96}.  
The spherically symmetric limb-brightened morphology, the low surface 
brightness and the low electron density of this shell \citep{GM96} 
are consistent with the expectations for an evolved object in the ISW 
model of PN formation \citep[e.g.,][]{Schonberner_etal10}.  
This conjecture is supported by the large kinematical age of 
12,500 yr derived from the simple comparison of its angular 
size, expansion velocity\footnote{
Even if we accounted for the fact that the real expansion velocity of 
the shock front is larger by 10\%--20\% than the observed velocity 
\citep{Schonberner_etal05b}, the kinematical age of A\,30 would still 
be large, $\sim$10,500 yr. }
\citep[38.5$\pm$1.0 km~s$^{-1}$, ][]{ML96} and distance 
\citep[1.7 kpc, ][]{CKS92}.  

Deep [O~{\sc iii}] images of A\,30 (Figure~\ref{img.opt}-{\it center}) 
reveal a different picture.  
The round nebular shell is filled by a delicate system of arc-like 
features that extend up to $\sim$30\arcsec\ from the central star 
and depict a cloverleaf pattern \citep{Jacoby79}.  
More remarkably, a series of knots are detected just a few 
arcsecs from the central star.  
These knots are resolved by \emph{HST} Wide Field Planetary Camera 2 (WFPC2) 
[O~{\sc iii}] images (Figure~\ref{img.opt}-{\it right}) to be distributed 
along a disk and two bipolar outflows \citep{Borkowski_etal95}, a physical 
structure later confirmed by the spatio-kinematical study carried out by 
\citet{CCC97}. 
The knots are found to be extremely faint in H~{\sc i} recombination 
lines\footnote{
The [N~{\sc ii}] $\lambda\lambda$6548,6583 and He~{\sc ii} $\lambda$6560 
to H$\alpha$ line ratios for the bipolar knots are $\sim$2.1, $\sim$6.3, 
and $\sim$2.4, respectively \citep{WLB03}.  
The emission detected at the position of these knots in the H$\alpha$ 
image (Figure~\ref{img.opt}-{\it left}) consists mostly of [N~{\sc ii}] 
and He~{\sc ii} emission, rather than H$\alpha$ emission itself, 
given the 60 \AA\ bandwidth of the H$\alpha$ filter used for these 
observations.  
}
\citep{WLB03} implying low content of hydrogen with a He/H abundance 
ratio of 10.8--11.6 and very high metal abundances of C/H$\sim$0.45, 
N/H$\sim$0.30 and O/H$\sim$1.30 \citep{JF83,GM96,WLB03,Ercolano_etal03}.   
These knots and the central star of A\,30 are embedded in large amounts 
of circumstellar dust \citep{Borkowski_etal94} with anomalous carbonaceous 
composition \citep{Greenstein81}.  
Apparently, these hydrogen-depleted knots have been ejected only 
recently by the 
central star of A\,30, in contrast to the old and hydrogen-rich 
outermost round shell.

The hydrogen-deficient nature of the CSPN of A\,30 and the presence of 
hydrogen-poor ejecta near the star inspired the born-again PN scenario 
\citep{Iben_etal83} (also known as ``very late thermal pulse'' or VLTP) 
in which the thermonuclear burning of hydrogen in the remnant stellar 
envelope builds up the amount of helium until its fusion into carbon and 
oxygen is ignited \citep[see, e.g., ][for recent theoretical studies of this 
phenomenon]{Herwig_etal99,Althaus_etal05,LM06,MA06,Miller_etal06}.  
Since the remnant envelope is shallow, the increase of pressure 
from this last helium shell flash leads to the ejection of the 
newly processed material in the envelope, while the stellar 
structure remains intact. 
As the remnant envelope expands, the stellar effective temperature 
decreases and the star returns to the AGB phase.  
The stellar evolution that follows this event is fast and will take the 
star back toward the post-AGB track in the HR diagram \citep[see, e.g., 
Figures 5 and 8 in][]{Miller_etal06}: the envelope of the star contracts, 
its effective temperature increases and a new fast stellar wind develops.  
In a sense, the PN is born-again.

The fast stellar wind will blast the hydrogen-poor material ejected 
by the star during the born-again event and the subsequent born-again 
AGB phase.   
This interaction has been captured by \emph{HST} WFPC2 images of 
the vicinity of the central star of A\,30 \citep{Borkowski_etal95}.  
The hydrogen-poor knots of A\,30 display a cometary appearance with either 
bow-shock structures pointing toward the central star or compact cores 
with fanning tails pointing away from the central star 
(Figure~\ref{img.opt}-{\it right}) 
whose expansion velocities increase outward up to $\sim$200 km~s$^{-1}$ 
\citep{ML96,CCC97}.  
Soft X-ray emission from the mixture of shocked stellar wind and
evaporated material can be expected \citep{Borkowski_etal95}, and 
has been confirmed by \emph{ROSAT} PSPC serendipitous observations 
that revealed a source of soft X-ray emission at a plasma temperature 
$\sim$3$\times$10$^5$~K \citep{ChuHo95}.  
A follow-up \emph{ROSAT} HRI observation showed a central point
source and hints of diffuse emission associated with the innermost
hydrogen-poor knots, although the detection of the diffuse emission 
is uncertain due to the low S/N ratio \citep{CCC97}.

We have obtained \emph{Chandra} and \emph{XMM-Newton} observations 
of A\,30 in order to accurately determine the spatial and spectral 
properties of its X-ray emission.  
The results are analyzed in conjunction with the physical properties of its 
stellar wind determined from refined non-LTE model fits to optical and UV 
spectra. 
Multi-epoch \emph{HST} archival images have also been used to 
search for proper motions of the hydrogen-poor knots in order 
to assess their angular expansion rate and to investigate their 
interactions with the fast stellar wind.

In the following, we first investigate in Sect.\ 2 the stellar wind 
properties of the CSPN of A\,30, given the implications for the X-ray 
emission, and determine the proper motions of the hydrogen-poor knots 
in Sect.\ 3.  
The X-ray observations and the spatial and spectral properties of 
the X-ray emission are described in Sect.\ 4 and the results are 
discussed in Sect.\ 5.  
The conclusions are presented in Sect.\ 6.

\section{Non-LTE Analysis of the CSPN of A\,30}

Optical and UV spectra of the CSPN of A\,30 have been analyzed 
using calculations performed with the Potsdam Wolf-Rayet (PoWR) 
model atmosphere code \citep[][and references therein]{HG04}.  
This code solves the non-LTE radiative transfer for a spherically expanding 
atmosphere, accounting for complex model atoms and line blanketing, to derive 
basic stellar and wind parameters.  
The calculations applied here include He, C, N, O, Ne, and the elements 
of the iron-group (the latter in the superlevel approximation).

The UV spectra of the CSPN of A\,30 were observed by the \emph{Far 
Ultraviolet Spectroscopic Explorer} (\emph{FUSE}) and \emph{International 
Ultraviolet Explorer} (\emph{IUE}) satellites.  
Data from these observations have been retrieved from MAST, the 
Mikulski Archive for Space Telescopes at the Space Telescope Science 
Institute\footnote{
The Space Telescope Science Institute (STScI) is operated by the 
Association of Universities for Research in Astronomy, Inc., under 
NASA contract NAS5-26555.}.  
The \emph{FUSE} observations of A\,30 in the spectral range 920--1180 \AA\ 
consisted of the data set B0230101 obtained on 2001 April 10 with the LWRS 
aperture for a total useful exposure time of 4.1 ks (Guerrero \& De Marco, 
in preparation).  
Similarly, the \emph{IUE} observations of A\,30 in the spectral range 
1150--3200 \AA\ consisted of the data sets SWP07955LL and LWR06930LL 
obtained on 1980 February 15 with total exposure times of 1.5 and 3.0 
ks, respectively.  
Complementary high-dispersion optical spectra of the CSPN of A\,30 
were obtained using the Ultraviolet and Visual Echelle Spectrograph 
(UVES) on the 8m UT2 of the VLT at Paranal Observatory on 2003 
February 19  in the framework of the large project 167.D-0407 (PI: 
Napiwotzki).  
The observations consisted of two 300 s exposures that covered the 
spectral regions 3290--4525 \AA, 4605--5610 \AA\ and 5675-6642 \AA.

\begin{figure*}[btp]
\includegraphics[scale=0.85]{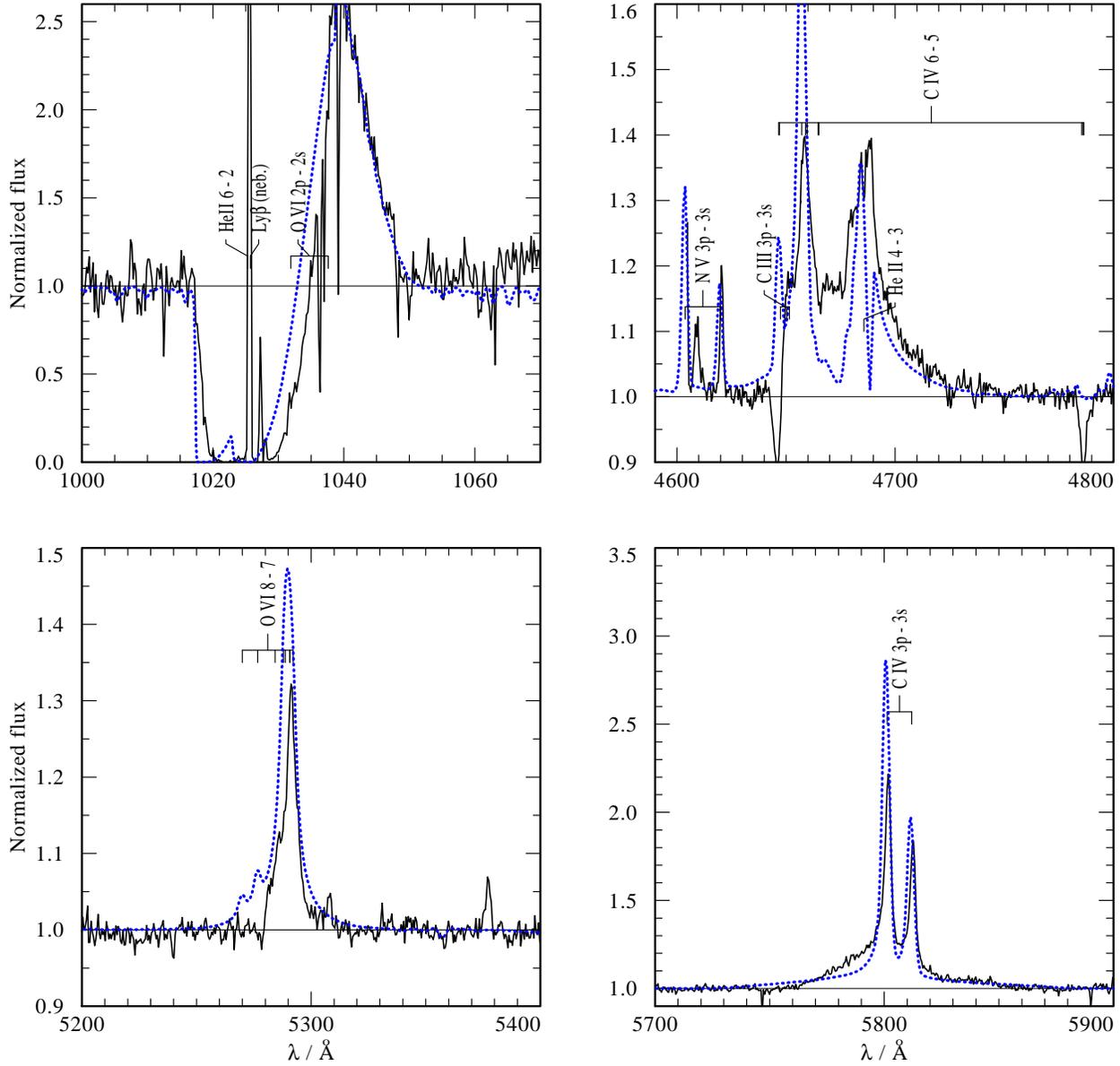}
\caption{Some of the spectral lines employed for the spectral analysis
of the central star of A\,30.  
Observations (solid line) are from \emph{FUSE} and from ground-based 
spectroscopy. 
The synthetic spectrum (dotted line) has been
calculated with the Potsdam Wolf-Rayet (PoWR) model atmosphere code,
using the parameters compiled in Table\,\ref{table:stellarparameters}.}
\label{fig:linefit}
\end{figure*}

In spite of the limited number of spectral lines useful for analysis 
provided by the UV and optical spectra of the CSPN of A\,30, a 
reasonable fit (see Fig.\,\ref{fig:linefit}) is achieved for the set 
of parameters compiled in Table\,\ref{table:stellarparameters}. 
For the fit we adopted a stellar luminosity of 6,000 $L_\odot$, 
noting that the stellar radius, mass-loss rate and distance 
scale with luminosity according to the relations shown in 
Table\,\ref{table:stellarparameters}.
The distance of 1.76 kpc, similar to the statistical distance 
of 1.69 kpc provided by \citet{CKS92}, will be used hereafter.  
The CSPN of A\,30 is confirmed to be very hot ($T_\ast = 115$\,kK), and the 
emission line spectrum originates from a stellar wind composed predominantly 
of helium, carbon and oxygen, which is typical for the [WC] spectral type.

Besides small changes in the values of the stellar temperature and 
helium and carbon abundances, the present results do not differ 
appreciably from those previously reported by \citet{Leuenhagen_etal93} 
based on a much earlier version of our model atmosphere code without 
the inclusion of iron-line blanketing.  
The major difference between the current and earlier calculations, however, 
is induced by the inclusion of clumping and mass-loss rate ($\dot{M}$) 
effects. 
In the ``microclumping'' approximation \citep[e.g.,][]{HK98}, the emission 
line fit yields the product $\dot{M}\sqrt{D}$ where $D$, the so-called 
``clumping factor'', is difficult to constrain.  
The current calculations adopt a value of 10 for $D$, which 
has been proven to be an adequate choice for massive WC stars.  
The only study for the wind of a CSPN \citep{THG08} came to a similar 
result, although this parameter was poorly constrained.  
A value of 10 for the clumping factor leads in the present calculations 
to a mass-loss rate ($\dot{M}=2\times10^{-8}\, M_\odot$~yr$^{-1}$) 2.5 
times lower than the value 
derived by \citet{Leuenhagen_etal93}.

According to our calculations, the stellar wind of the CSPN of A\,30 
has a He:C:N:O element number ratio 100:11:0.7:6.0, i.e., it is carbon 
rich with a C/O ratio $\sim$1.8.  
It is interesting to compare these chemical abundances with those of 
the hydrogen-deficient knots, where the He:C:N:O element number ratio 
is 100:4.0:2.7:11.6 \citep{WLB03,Ercolano_etal03}.  
The low C/O ratio of the hydrogen-poor knots, $\sim$0.3, is at 
variance with the stellar wind, and is also in contradiction to 
theoretical models of born-again PNe that predict C/O higher than 
unity \citep[e.g.,][]{Iben_etal83,Herwig_etal99,Miller_etal06}.

\begin{figure*}[t]
\includegraphics[width=2.15\columnwidth]{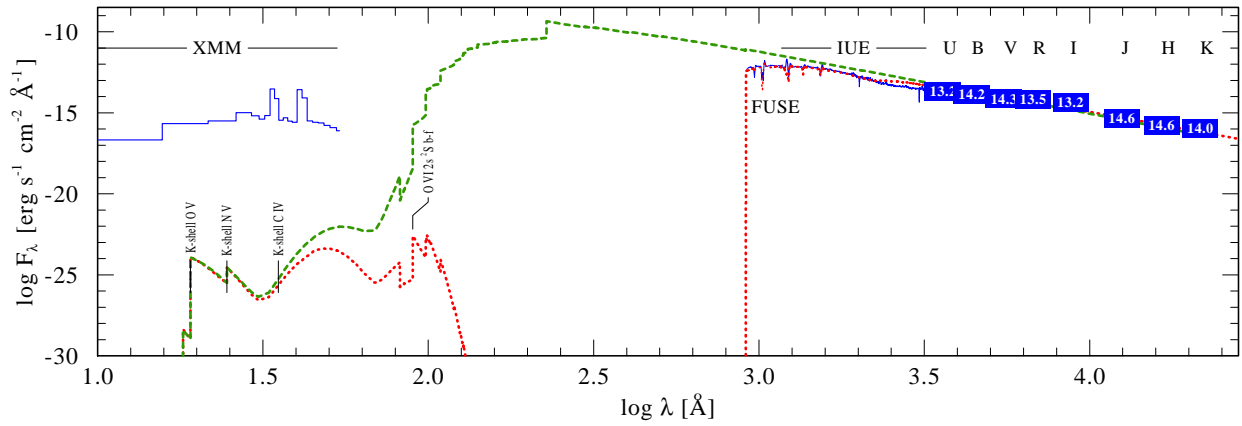}
\caption{
Spectral energy distribution (SED) of the central star of A\,30 from the 
infrared to X-ray range. 
Observations (blue) are photometric measurements in the indicated bands, 
calibrated UV spectra from the \emph{IUE} and \emph{FUSE} satellites, and 
the \emph{XMM-Newton} observations reported in this paper. 
These measurements are compared to the theoretical SED from our stellar 
model with the parameters compiled in Table\,\ref{table:stellarparameters}.  
The green dashed line shows the SED without interstellar reddening,
and the red dotted line after interstellar extinction has been applied. 
}
\label{fig:sed}
\end{figure*}

Finally, the spectral fit of the UV and optical spectra of A\,30 
(Fig.\,\ref{fig:linefit}) has allowed us to build its spectral 
energy distribution (SED) shown in Figure\,\ref{fig:sed}. 
The fit of the SED shown in Fig.\,\ref{fig:sed} requires only a 
small interstellar reddening with $E_{B-V}$=0.18$\pm0.05$\,mag 
that corresponds to a hydrogen column density 
$N_{\rm H}$=(6.8$\pm$1.9)$\times$10$^{20}$ cm$^{-2}$ according to 
the relation $N_{\rm H}$/$E_{B-V}$=3.8$\times$10$^{21}$ 
cm$^{-2}$~mag$^{-1}$ prescribed by \citet{GL89}.  
We note that the interstellar extinction law of \citet{CCM89} used 
to deredden the observed spectral data does not reproduce properly 
the UV absorption at $\sim$2470 \AA\ which is attributed to 
carbonaceous dust \citep{Greenstein81}, and thus the value of the 
hydrogen column density given above is suspect.

\section{Proper Motions of the H-poor Knots of A\,30}

\begin{figure*}
\includegraphics[width=6.5in]{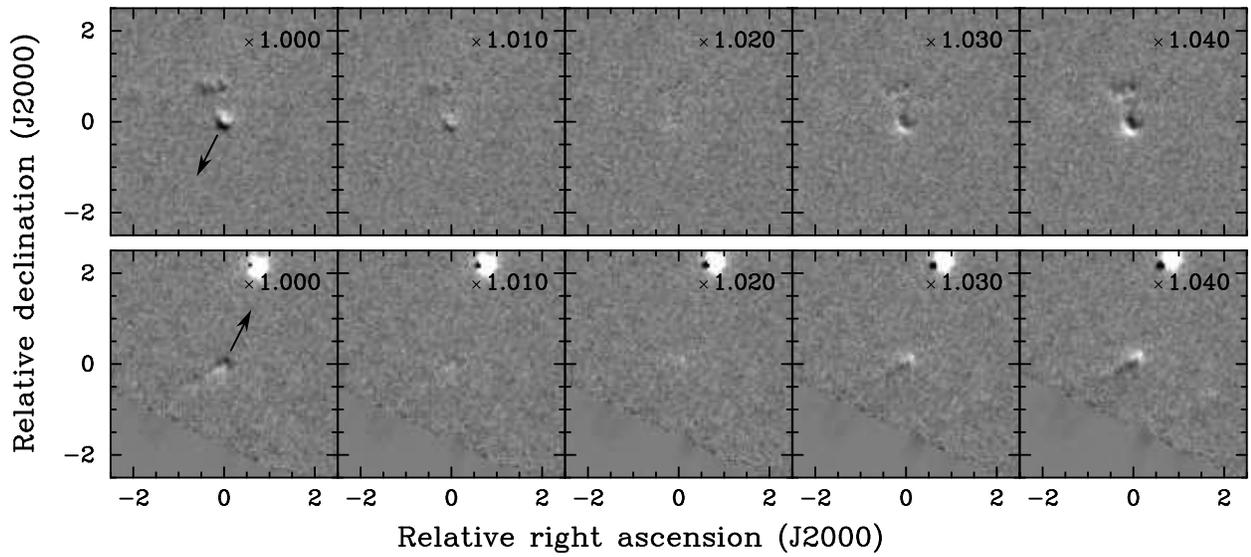}
\caption{
Residual maps of the northern (upper panels) and southern (lower panels) 
hydrogen-poor bipolar knots of A\,30, where the F502N image has been 
magnified by the factor noted in each panel.  
Dark shades correspond to the F502N March 1994 image and bright 
shades to the F555W December 2009 image.  
The direction of the central star is marked by an arrow in the leftmost 
panels. 
}
\label{fig:exp}
\end{figure*}

The original \emph{HST} WFPC2 narrow-band [O~{\sc iii}] images of A\,30 
(Fig.~\ref{img.opt}-{\it right}) were obtained on 1994 March 6 (epoch 
1994.2), but we noticed that the \emph{HST} archive also contained Wide 
Field Camera 3 (WFC3) images obtained through the F555W filter on 2009 
December 31 (epoch 2010.0).  
It is reasonable to compare these images, taken $\sim$15.8 yr apart, 
because the nebular emission registered by the F555W filter is mostly 
dominated by the [O~{\sc iii}] emission lines.  
Such comparison indeed unveils the proper motion of the hydrogen-poor 
knots of A\,30, as shown in Figure~\ref{fig:exp}. 
Following the method used by \citet{Reed_etal99} to study the angular 
expansion of NGC\,6543, we have magnified the earlier epoch image (the 
F502N image) by several factors and produced the residual maps shown 
in Fig.~\ref{fig:exp}.  
These maps suggest that the hydrogen-poor knots of A\,30 have 
expanded $\sim$2\%.  

A detailed analysis of the location of both bipolar and equatorial knots 
along different directions in the images of the two epochs has allowed 
us to refine this result and conclude that the 1994.2 image needs to be 
magnified by 1.019$\pm$0.003 in order to match the 2010.0 image.  
Such 1.9\% expansion in a time lapse of 15.8 yr implies an expansion 
age of 850$^{+280}_{-150}$ yr that can be interpreted as the time since 
the born-again episode took place circa {\sc ad} 1160.

The angular expansion rate can also be used to estimate the averaged 
expansion velocity of the knots in the equatorial ring.  
Assuming that the equatorial ring is circular in shape, its major axis lies 
on the plane of the sky and thus the semimajor axis of 4\farcs8 implies a 
radius of 0.041 pc at a distance of 1.76 kpc.  
For an age of 850 yr, an averaged expansion velocity 
of $\sim$50 km~s$^{-1}$ is derived.  

%

Finally, we note the presence of a star $\sim$5\farcs25 from the CSPN 
of A\,30 at PA$\sim$144\arcdeg.  
The possible physical connection of this star with the CSPN of A\,30 
was used by \citet{Ciardullo_etal99} to estimate a distance of 
2020~pc toward A\,30.  
The comparison between the 1994.2 and 2010.0 epoch images reveals 
a change of 0\farcs15 in the position of this star relative to the 
location of the CSPN of A\,30.  
This shift is much larger than the orbital motion expected in the time 
lapse of 15.8 yr for a companion star with the orbital separation of 
10,580 AU estimated by \citet{Ciardullo_etal99}. 
We therefore conclude that this star and the CSPN of A\,30 do not form 
a binary system, but they are rather optical doubles.  
Consequently, the distance estimate of 2020 pc is not valid and should 
not be used.

\section{X-ray Observations of A\,30}


A\,30 was observed by \emph{XMM-Newton} on 2009 October 21 (Observation 
ID 0605360101, PI: W.-R.\ Hamann) using the European Photon Imaging 
Camera (EPIC) and the Reflection Grating Spectrometer (RGS) instruments  
for a total exposure time of 58.2 ks.  
The EPIC observations were performed in Full Frame Mode with the 
Thin Filter.   
The second version of the observation data files (ODFs) generated 
by the \emph{XMM-Newton} Science Operation Center on 2010 June 18 
were processed using the \emph{XMM-Newton} Science Analysis Software 
(SAS) 10.0.2.   
Reprocessed EPIC-MOS, EPIC-pn and RGS event lists were created 
using the SAS tools ``emproc'', ``epproc'' and ``rgsproc'', 
respectively, and the most up-to-date \emph{XMM-Newton} calibration 
files available on the Current Calibration File (CCF) as of 2010 
September 15.  
The original exposure times were 40.88, 41.08, 37.23, 57.43 and 57.47 ks 
for EPIC-MOS1, EPIC-MOS2, EPIC-pn, RGS1 and RGS2, respectively, but the 
last segment of the observations was dramatically affected by periods of 
high background.  
After excising these periods from the data, the resulting useful 
exposure times amount to 31.63 ks for EPIC-MOS1, 31.31 ks for 
EPIC-MOS2, 24.58 ks for EPIC-pn, 31.81 ks for RGS1 and 31.53 ks 
for RGS2. 

A\,30 was subsequently observed by \emph{Chandra} on 2011 January 1 
(Observation ID 12385, PI: Y.-H.\ Chu) using the array for spectroscopy 
of the Advanced CCD Imaging Spectrometer (ACIS-S) for a total exposure 
time of 96.09 ks.  
The nebula was imaged on the back-illuminated CCD S3 using the VFAINT mode.  
No periods of high background affected the data and the resulting 
useful exposure time amounts to 96.08 ks after excising dead-time 
periods.  
The \emph{Chandra} Interactive Analysis of Observations (CIAO) software 
package version 4.3 was used to analyze these data.

The \emph{XMM-Newton} EPIC observations detect a relatively bright source 
at the location of the central star of A\,30.  
An inspection of the images at different energy bands shown in 
Figure~\ref{fig:ximg} indicates that this source is soft, with 
emission from the lowest energies detectable by the EPIC cameras 
up to 0.6 keV, above which little or no emission is seen.  
%
%
Similarly, the \emph{Chandra} observations detected a soft source at 
the position of the CSPN of A\,30. 
%
The EPIC-pn, EPIC-MOS and ACIS-S background-subtracted count rates and 
net number of counts detected in different energy ranges are provided 
in Table~\ref{table:counts}.

\subsection{Spatial Properties of the X-ray Emission from A\,30}

\begin{figure*}[t]
\includegraphics[width=6.5in]{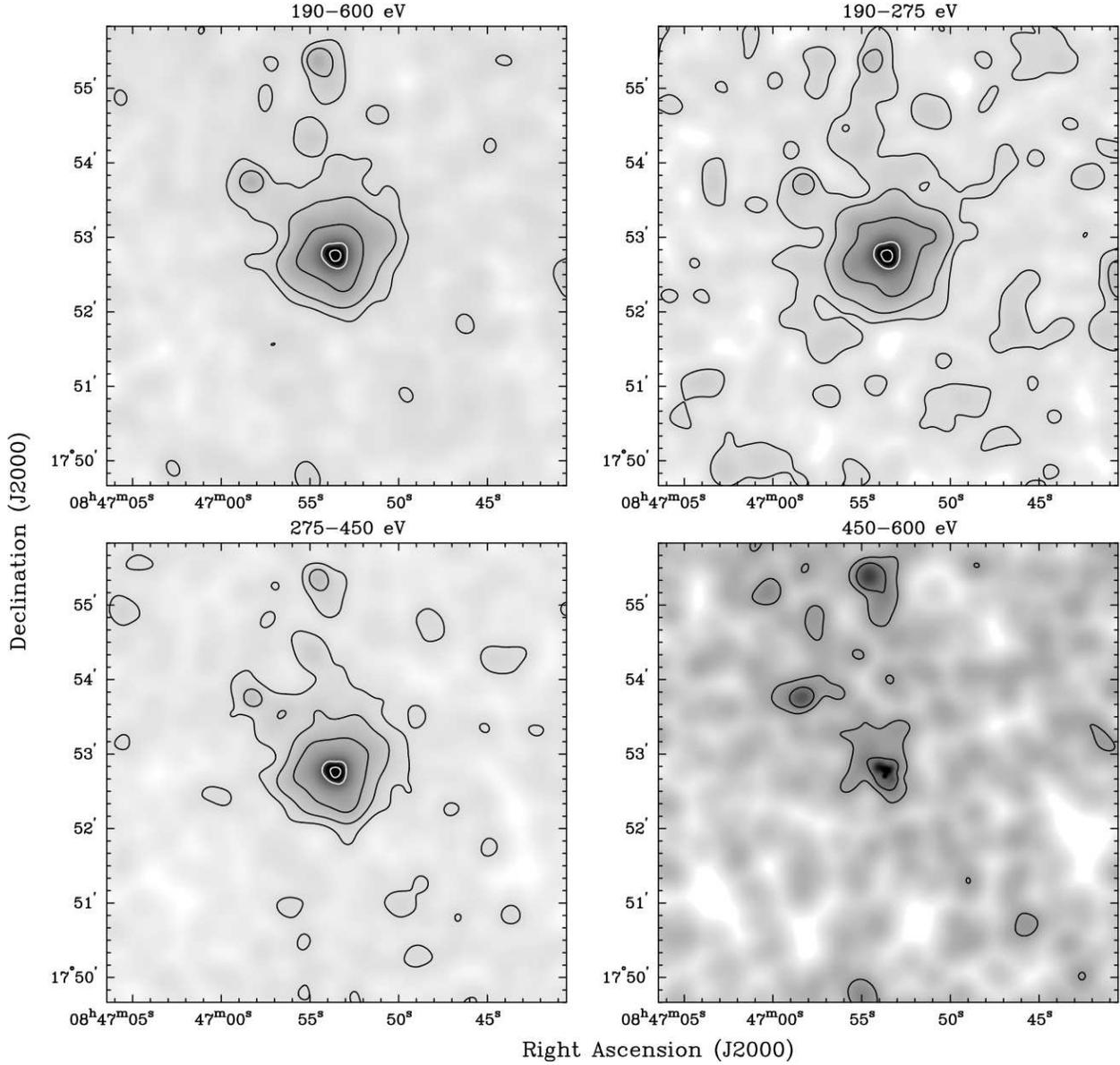}
\caption{
Exposure-corrected \emph{XMM-Newton} EPIC images of A\,30 in 
different energy bands. 
The images have a pixel size 1\arcsec\ and have been smoothed using an 
adaptive Gaussian kernel with sizes between 1\arcsec\ and 8\arcsec.
Gray-scales have been chosen in the range between 15$\sigma$ below 
the background level and 20\% of the intensity peak for each image. 
The black lower contours correspond to 5$\sigma$, 10$\sigma$, 
and 30$\sigma$ over the background level, while the white upper 
contours represent 10\% and 50\% of the peak intensity.  
}
\label{fig:ximg}
\end{figure*}

\begin{figure*}
\includegraphics[width=3.0in]{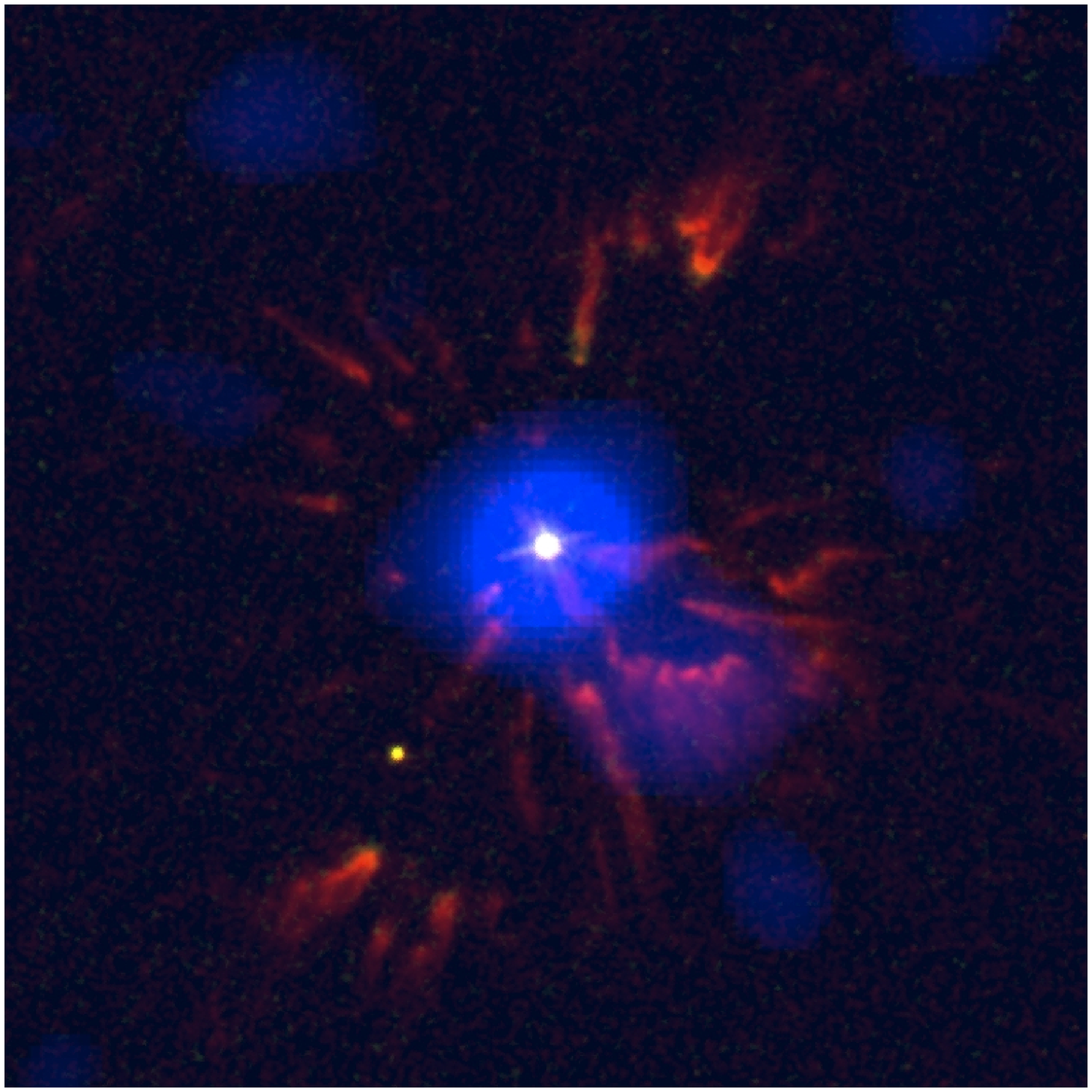}
\includegraphics[width=3.0in]{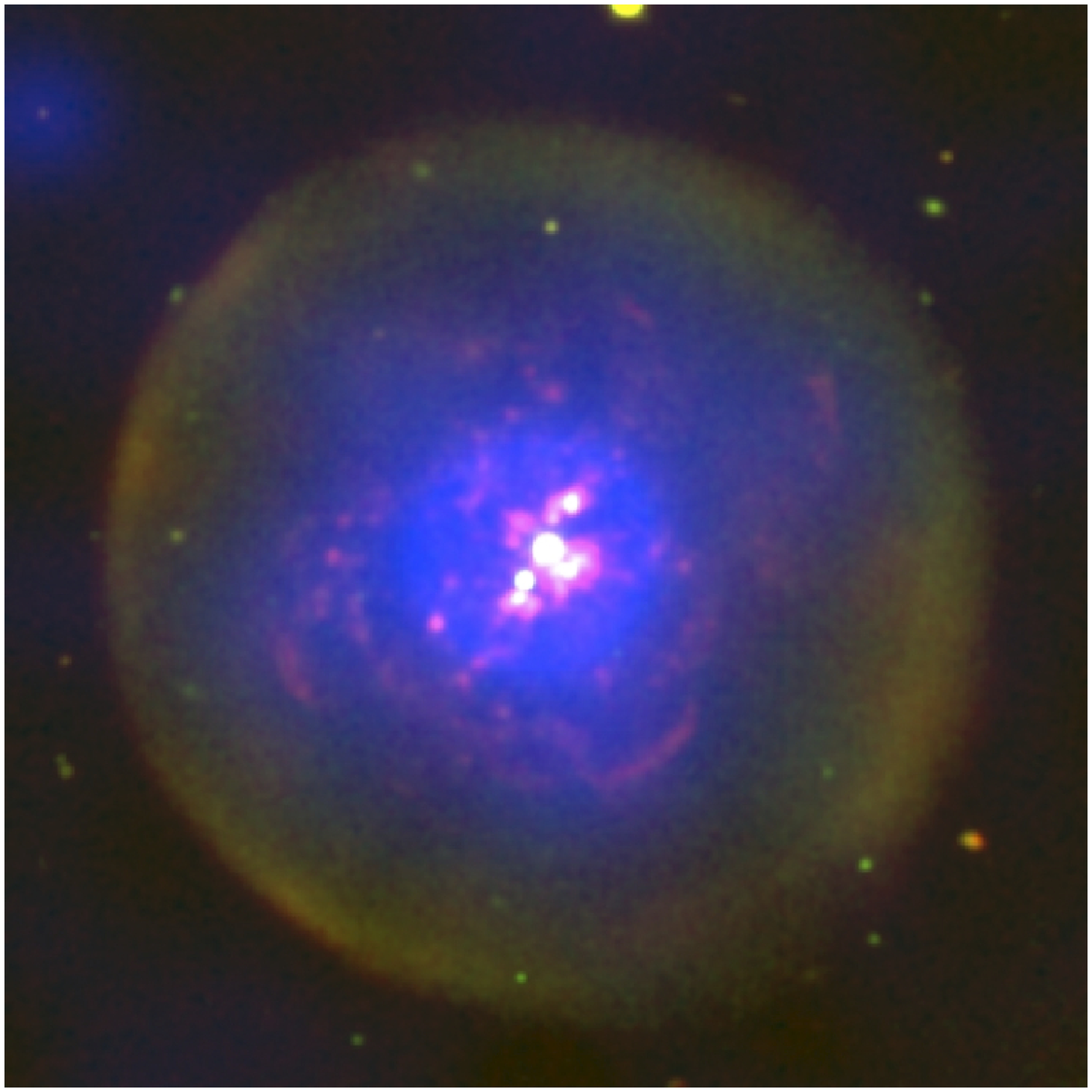}
\includegraphics[angle=270,width=6.5in]{A30_xopt.eps}
\caption{
Comparison of narrow-band optical and X-ray images of A\,30.  
({\it top}) 
Color-composite images: 
the \emph{HST} WFPC2 (RED=[O~{\sc iii}], GREEN=He~{\sc ii}) and 
\emph{Chandra} ACIS-S 0.20-0.60 keV (BLUE) composite in the left panel 
illustrates the small-scale spatial distribution of X-rays, whereas 
their large-scale spatial distribution is shown by the ground-based 
(RED=[O~{\sc iii}], GREEN=H$\alpha$) and \emph{XMM-Newton} 0.19-0.60 
keV (BLUE) composite in the right panel.  
({\it bottom}) 
Optical [O~{\sc iii}] narrow-band images of A\,30 overplotted by X-ray 
contours: 
\emph{HST} WFPC2 image overplotted by \emph{Chandra} ACIS-S (blue and 
black) and \emph{ROSAT} HRI (red) contours ({\it left}) and ground-based 
image overplotted by \emph{XMM-Newton} EPIC (blue and black) and 
\emph{ROSAT} PSPC (red) contours 
({\it right}).  
Red and blue contours have been set at 95\%, 75\%, 50\%, and 25\% of 
the peak intensity, whereas the black contours correspond to 5$\sigma$, 
10$\sigma$, 25$\sigma$, and 100$\sigma$ above the background level for 
the \emph{Chandra} ACIS-S image, and 10$\sigma$, 25$\sigma$, 50$\sigma$, 
and 100$\sigma$ above the background level for the \emph{XMM-Newton} 
EPIC image.  
}
\label{img.xopt}
\end{figure*}

In order to study the spatial distribution of the X-ray emission from 
A\,30, we have produced EPIC images of A\,30 in different energy bands 
(Figure~\ref{fig:ximg}) by extracting the individual EPIC-pn, EPIC-MOS1 
and EPIC-MOS2 images, mosaicing them together, applying the exposure 
map correction, and smoothing the images.  
We have also produced a \emph{Chandra} ACIS image in the 200-600 eV 
energy band.  
The \emph{Chandra} and \emph{XMM-Newton} X-ray images are compared 
to optical narrow-band images and previous \emph{ROSAT} X-ray images 
in Figure~\ref{img.xopt}.

The small-scale spatial distribution of the X-ray emission is revealed by 
the comparison between the \emph{Chandra} ACIS-S and \emph{HST} WFPC2 
[O~{\sc iii}] images shown in Figure~\ref{img.xopt}-{\it left}.  
Most of the emission detected in the \emph{Chandra} ACIS-S image 
corresponds to a point source coincident with the central star of 
A\,30, but some additional emission, 8.2$\pm$3.5 counts, is found 
$\sim$4\arcsec\ to the southwest of A\,30 CSPN.  
This emission is soft, with a median energy $\sim$0.30 keV, 
and seems spatially coincident with an [O~{\sc iii}] bright 
knot in the equatorial ring of the hydrogen-poor ejecta.  
Given the extremely low count level of the ACIS-S background in the soft  
energy band, $\sim$0.025 counts~arcsec$^{-2}$, the probability that this 
source were a statistical fluctuation in the background is negligible, 
$<$10$^{-6}$.  
Furthermore, Figure~\ref{img.xopt} shows that this source  is coincident 
with one of the brightest patches of diffuse X-ray emission suggested by 
\emph{ROSAT} HRI images \citep{CCC97}.

The large-scale spatial distribution of the X-ray emission is 
illustrated by the comparison between the \emph{XMM-Newton} 
EPIC and ground-based [O~{\sc iii}] images shown in 
Figures~\ref{fig:ximg} and \ref{img.xopt}-{\it right}.  
The X-ray emission in the \emph{XMM-Newton} EPIC image peaks at the 
location of the central star, in agreement with the \emph{Chandra} 
ACIS-S image.  
The image does not show any obvious diffuse X-ray emission associated 
with the round outer shell of A\,30.  
Instead, the X-ray peak at the central star is surrounded by a ``halo'' 
of diffuse emission whose spatial distribution is consistent with that 
revealed by \emph{ROSAT} PSPC observations at a poorer spatial resolution 
\citep{ChuHo95}.  
There is a tantalizing correlation between the \emph{XMM-Newton} EPIC 
X-ray contours of this diffuse X-ray emission and the ``petals of the 
cloverleaf'' pattern interior to the round outer shell.

To further assess whether this X-ray emission is extended, we have used 
the SAS 10.0 task ``eradial'' to extract a radial profile of the X-ray 
emission from A\,30 and fit it to the theoretical \emph{XMM-Newton} 
EPIC-pn PSF that can be described as a King function with core 5\farcs5 and 
exponent 1.6.
%
%
However, a direct fit to the radial profile has yielded inconclusive results 
because A\,30 is located at $\sim$70\arcsec\ from a chip gap in the EPIC-pn 
camera, and thus does not allow the extraction of a sufficiently extended 
radial profile to accurately assess the background level for the PSF fit.  
The EPIC-MOS images are not useful for this purpose as they lack 
sufficient statistical significance.  

\begin{figure*}
\includegraphics[angle=0,scale=0.475,bb=60 235 552 575]{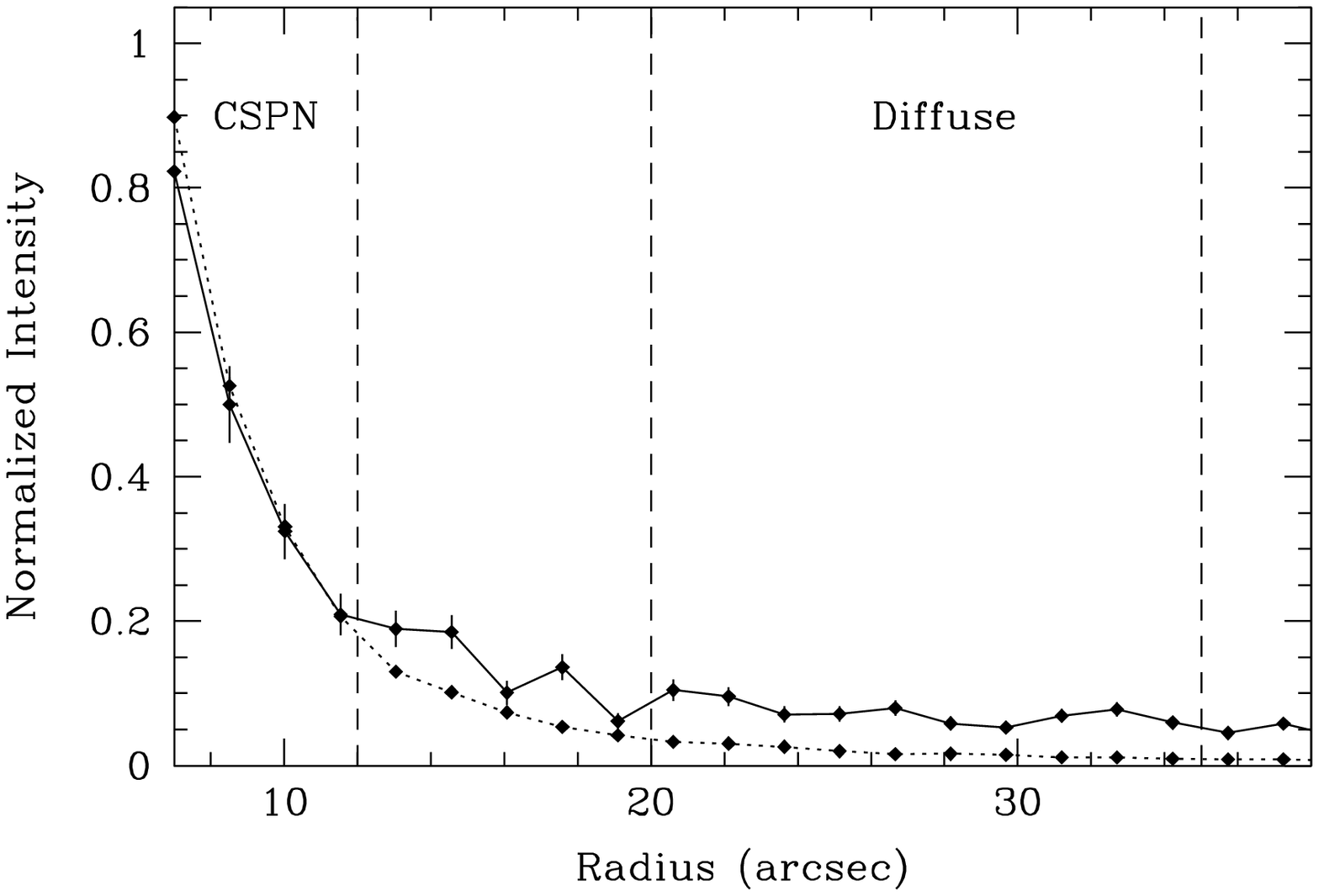}
\includegraphics[angle=0,scale=0.475,bb=60 235 552 575]{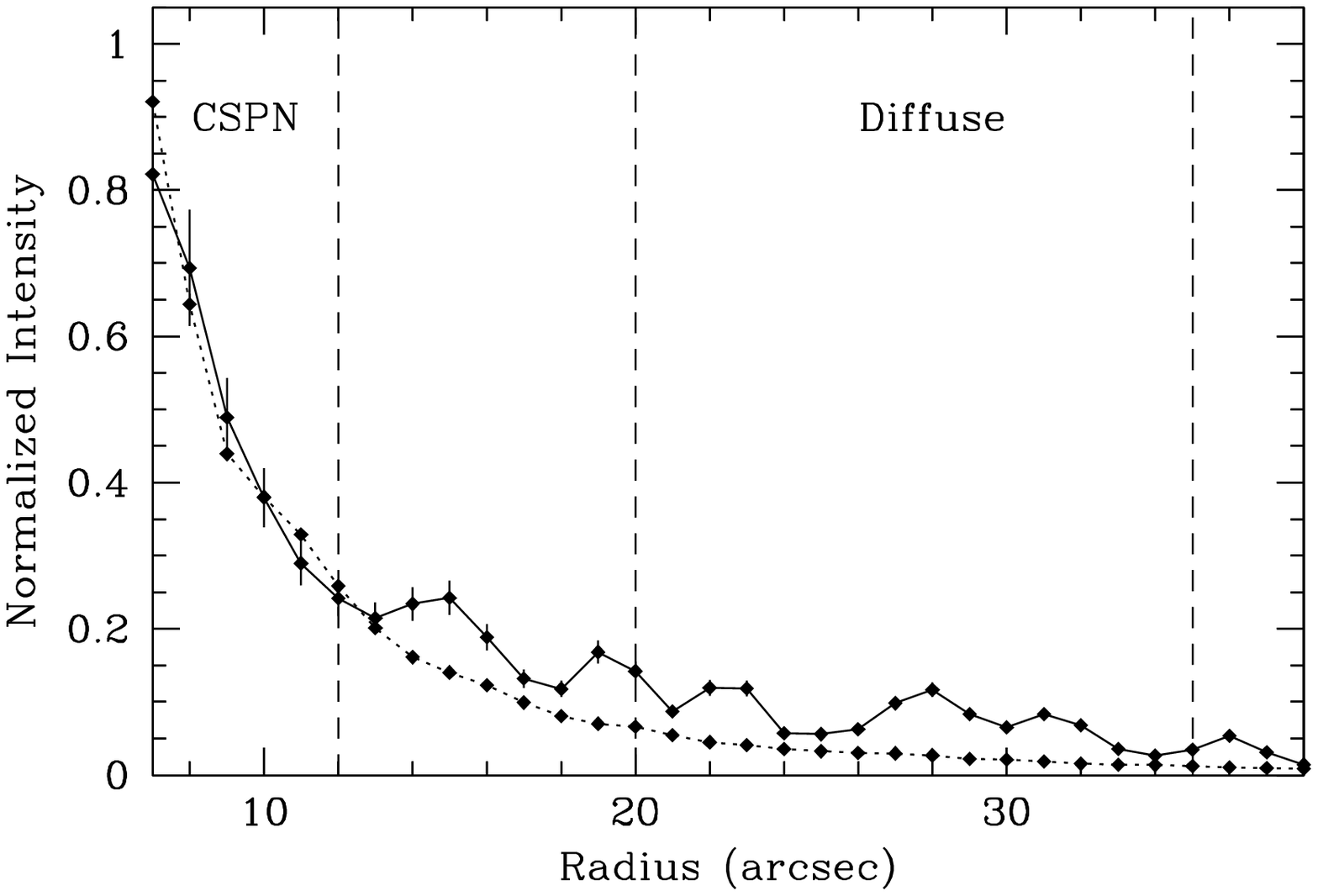}
\caption{
Comparison of the EPIC-pn radial profiles of A\,30 (solid line) and Nova 
LMC1995 (dotted line) as extracted using ``eradial'' {\it (left)} and 
by deriving the count rate in annular regions {\it (right)}.   
The vertical dashed lines mark the source regions for the extraction 
of the X-ray spectra of the CSPN and diffuse emission.  
}
\label{psf}
\end{figure*}

To overcome these difficulties, we have compared a bright, soft point 
source \citep[Nova LMC1995,][]{Orio_etal03} with A\,30. 
A preliminary inspection of the soft images of A\,30 and Nova LMC1995 
is also inconclusive because Nova LMC1995 also seems to be surrounded 
by a halo of diffuse emission most likely associated with the PSF of the 
EPIC-pn.  
A close comparison of the radial profiles of A\,30 and Nova LMC1995 
(Figure~\ref{psf}-{\it left}) built using the SAS task ``eradial'' 
finally found evidence suggesting that the radial profile of A\,30 
departs from that of a point source at distances $>$13\arcsec.
To reinforce this result, we have also determined the count rate in 
circular annuli centered on A\,30 and Nova LMC1995.  
The comparison, shown in Figure~\ref{psf}-{\it right}, confirms that 
A\,30 shows additional emission peaks at distances $>$13\arcsec, 
further strengthening the conclusion that A\,30 displays extended 
X-ray emission.  
The limited spatial resolution of the \emph{XMM-Newton} EPIC-pn 
observations, however, makes it impossible to estimate the extent 
and distribution of this diffuse emission within 20\arcsec\ from 
the central star of A\,30.  
An attempt to remove the emission from the point source has been made by 
using a point source model derived from the observation of Nova LMC1995 
and scaled to the emission peak of the CSPN of A\,30.  
The residual diffuse emission is basically consistent with 
the contours shown in Figure~\ref{img.xopt}-{\it right}.


\subsection{Spectral Properties of the X-ray Emission from A\,30}

\begin{figure*}
\includegraphics[angle=0,scale=0.40,bb=42 160 558 595]{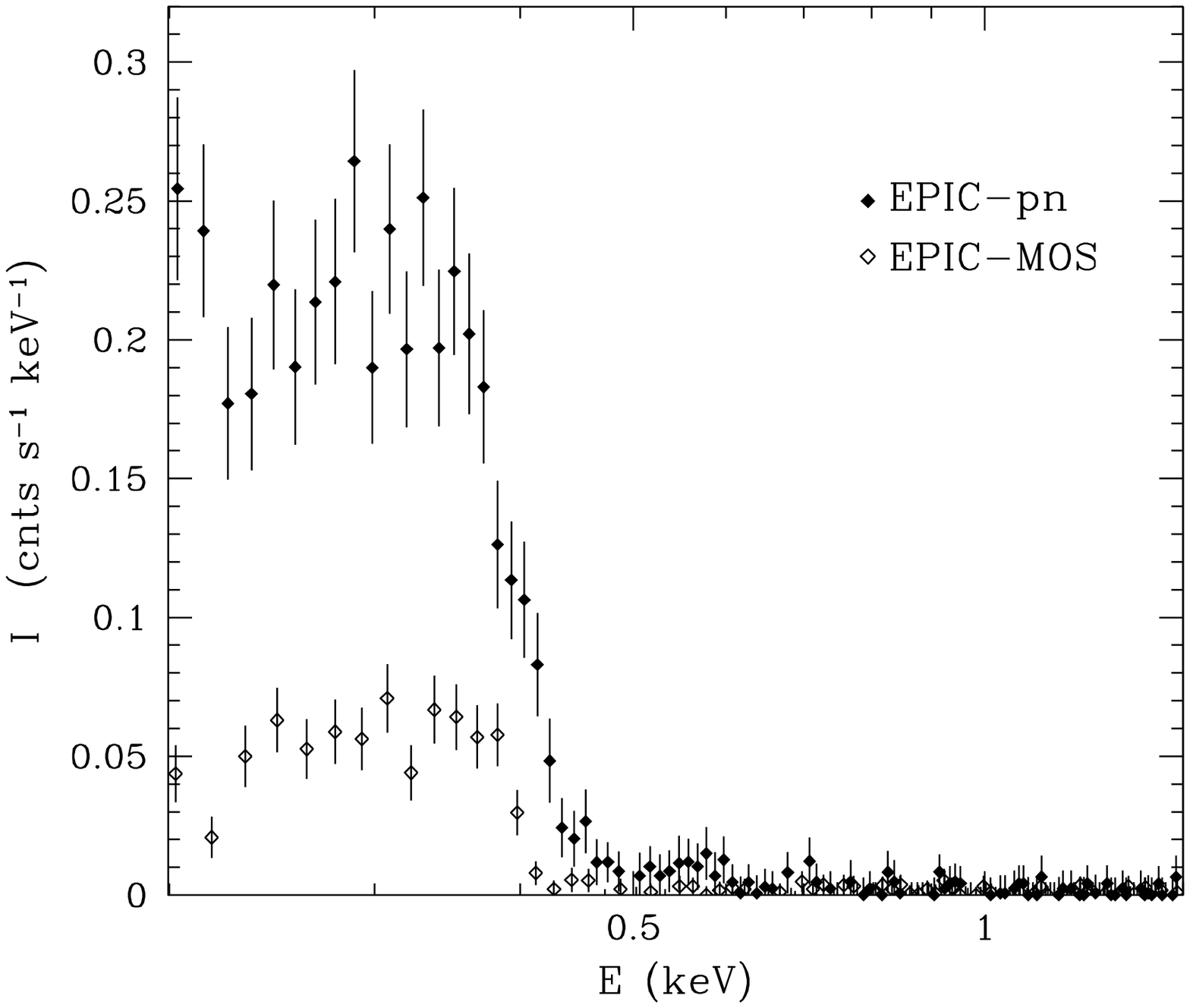}
\includegraphics[angle=0,scale=0.485,bb=35 235 585 700]{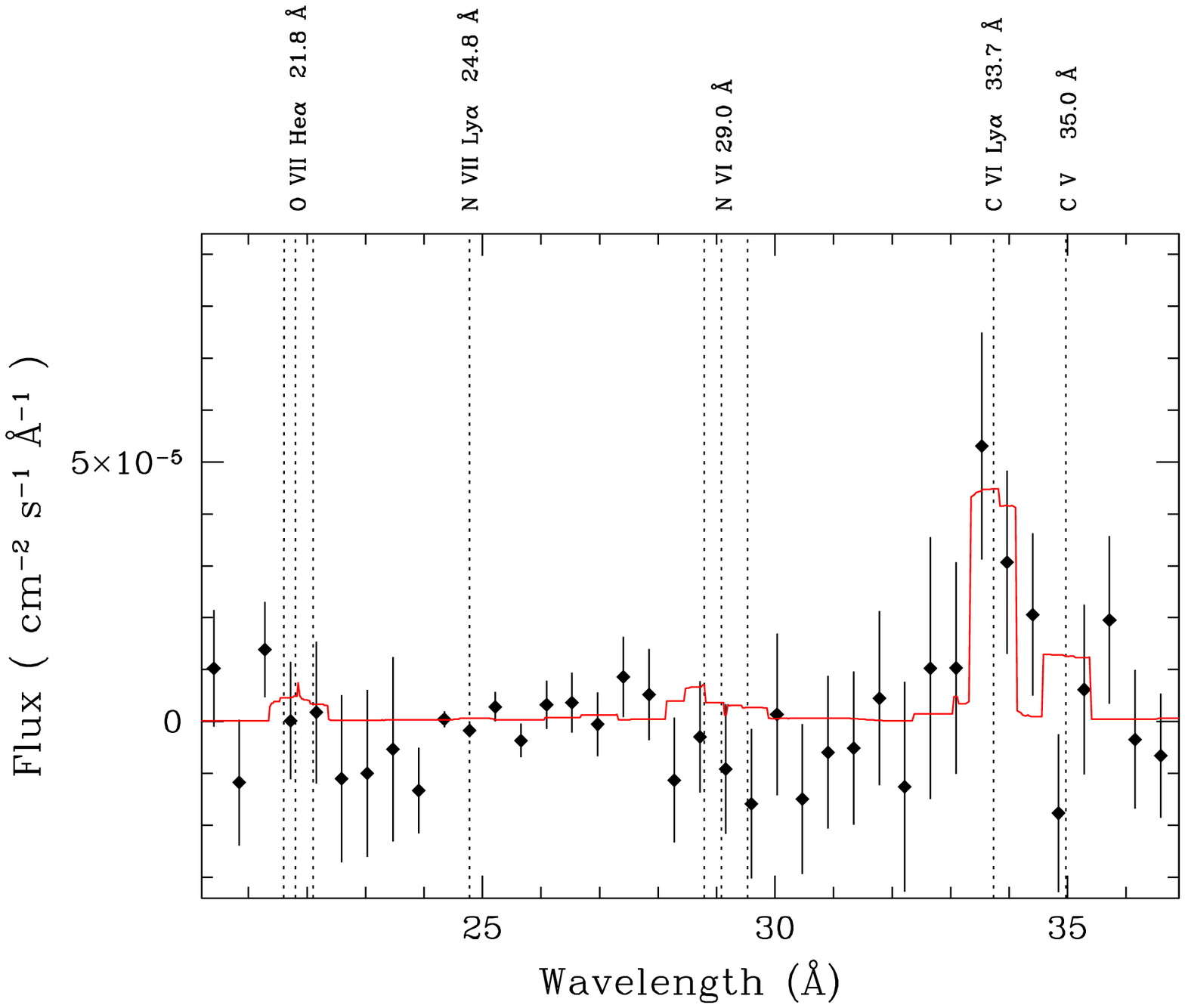}
\caption{
\emph{XMM-Newton} EPIC {\it (left)} and combined RGS1+RGS2 {\it (right)} 
background-subtracted spectra of A\,30.  
The EPIC spectra have been extracted from a circular aperture of 
radius 35\arcsec\ centered on A\,30, while the RGS spectrum has been 
extracted using the standard RGS aperture. 
The red histogram overplotted on the {\it (right)} panel corresponds 
to the plasma emission model described in the text at a similar spectral 
resolution as that of the RGS spectrum.  
}
\label{xspec_all}
\end{figure*}

To study the spectral properties of the X-ray emission from A\,30, we 
have extracted the EPIC-pn and EPIC-MOS background-subtracted spectra 
of A\,30 shown in Figure~\ref{xspec_all}-{\it left}.  
The spectra are extremely soft even when compared to those of 
diffuse emission from other PNe \citep[e.g., NGC\,6543,][]{Chu_etal01}.  
The EPIC-pn spectrum, which has the best signal-to-noise ratio, 
peaks at $\sim$0.35 keV with a shoulder or slow decline toward 
lower energies and a rapid drop in the energy interval from 0.35 
keV to 0.5 keV.  
There is much fainter emission at $\sim$0.58 keV, but no further 
emission is detected above 0.6 keV.  
A comparison with optically thin plasma emission models of different 
chemical abundances suggests that the emission at 0.35 keV may 
correspond to either the C~{\sc vi} lines at 33.7 \AA\ ($\equiv$0.37 
keV) or the C~{\sc v} lines at 35.0 \AA\ ($\equiv$0.35 keV), while 
the weak feature at 0.58 keV seems consistent with the O~{\sc vii} 
triplet at 21.8 \AA\ ($\equiv$0.57 keV).  
Similarly, the rapid decline above 0.35 keV seems consistent 
with plasma emission models for which the contribution from 
the N~{\sc vi} 0.43 keV and N~{\sc vii} 0.50 keV emission lines 
is rather small.  
The RGS spectrum of A\,30, despite having a limited signal-to-noise ratio 
(Figure~\ref{xspec_all}-{\it right}), has allowed us to identify the 
emission peak in the EPIC-pn spectrum of A\,30 as the Ly\,$\alpha$ line of 
C~{\sc vi} at 33.7 \AA\ ($\equiv$0.37 keV) and to confirm that there is no 
significant contribution from nitrogen lines.

\begin{figure*}
\includegraphics[angle=0,scale=0.45,bb=45 162 555 593]{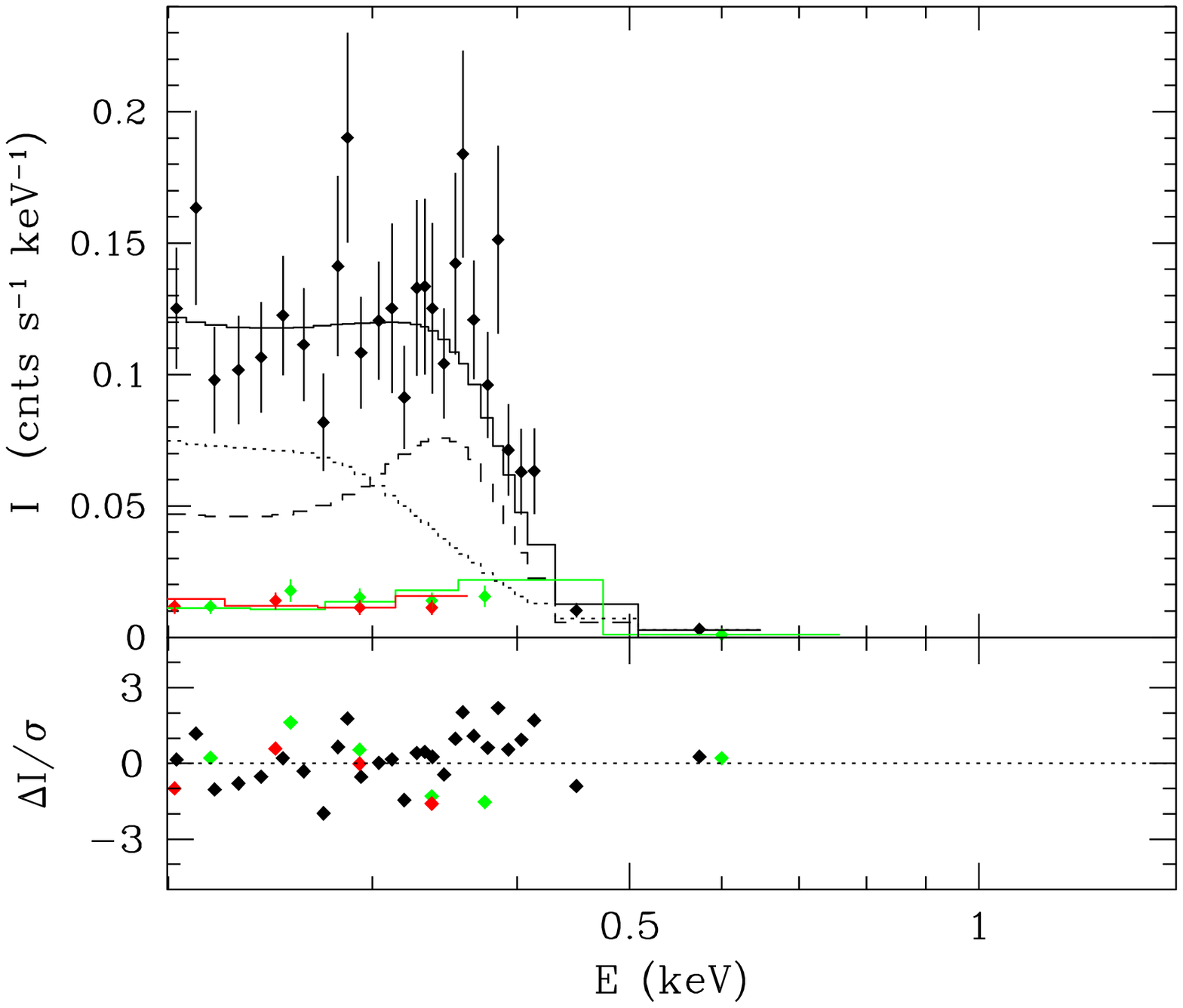}
\includegraphics[angle=0,scale=0.45,bb=45 162 555 593]{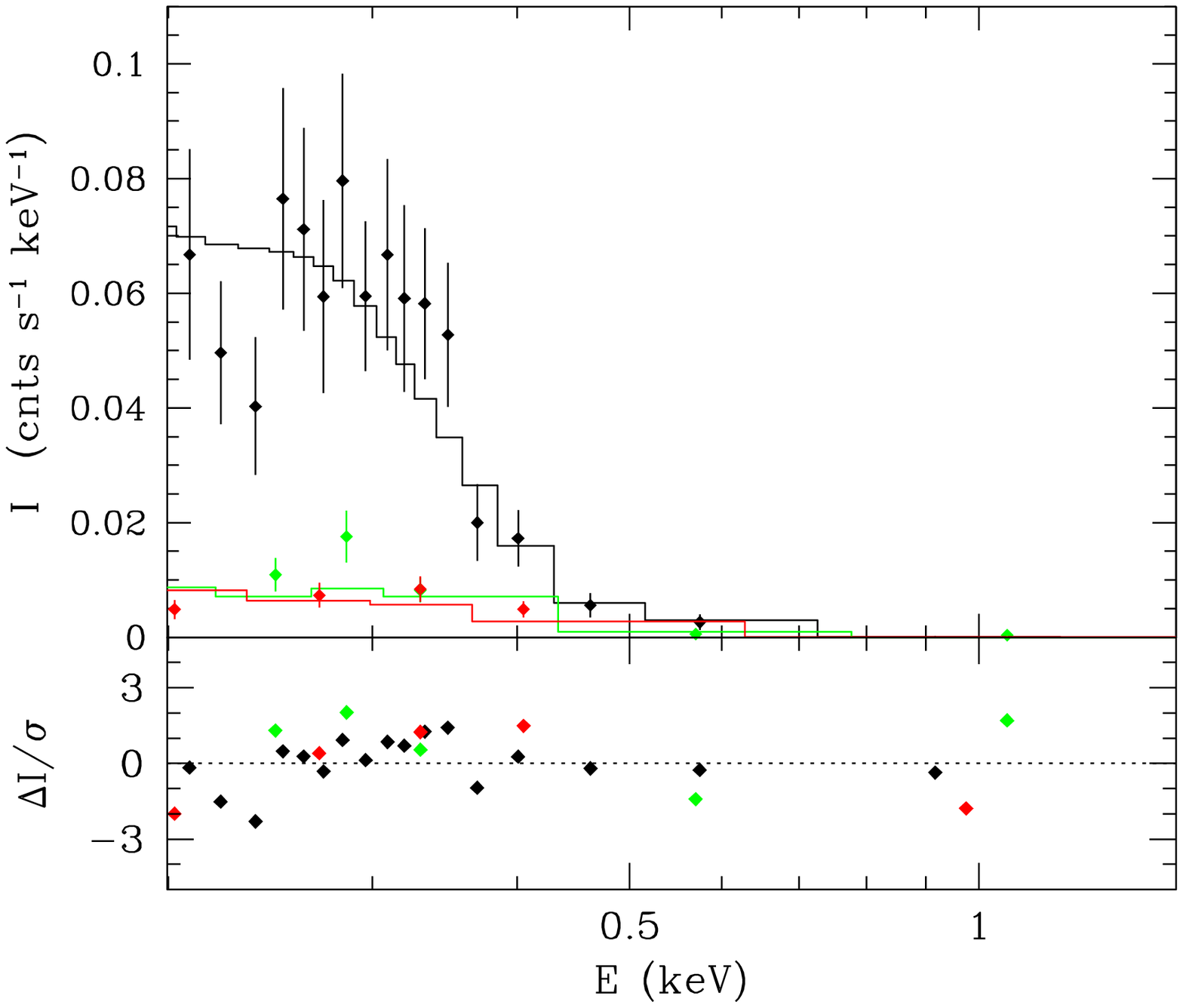}
\caption{
EPIC-pn (black), EPIC-MOS1 (green), and EPIC-MOS2 (red) 
background-subtracted spectra of the central source in A\,30 
{\it (left)} and its diffuse emission {\it (right)}.  
Note the different intensity scale as the spectrum of the diffuse 
emission is $\sim$2.5 times fainter than that of the central source.  
The solid lines correspond to the best-fit described in the text, 
with the residuals of the fit shown in the lower panels.  
For the spectral fit, the channels have been binned to include 20--25 
counts per channel.  
The dotted and dashed lines in the panel of the central star 
correspond to the thermal emission plasma and emission line 
components of the EPIC-pn model, respectively.  
}
\label{2xspec}
\end{figure*}

Since the analysis of the radial profile of X-ray emission from A\,30 
reveals extended emission, we have extracted separate spectra for the 
central source from a circular region of radius 12\arcsec\ and for the 
diffuse emission from an annular region with a 20\arcsec\ inner radius 
and a 35\arcsec\ outer radius.  
The background-subtracted EPIC-pn and EPIC-MOS spectra of the 
central source and diffuse emission are shown in Figure~\ref{2xspec} 
and their count rates and count numbers are listed in 
Table~\ref{table:counts}.   
The comparison of the spectra of the diffuse emission and point source 
suggests spectral differences, with the diffuse emission spectrum 
lacking the peak at $\sim$0.37 keV associated with the C~{\sc vi} line 
and having a relatively more important contribution of the O~{\sc vii} 
triplet at 0.57 keV.

\subsection{Spectral Analysis}

\subsubsection{X-ray Emission Model for A\,30}

The EPIC and RGS spectra of A\,30 imply the presence of emission lines, 
thus suggesting that the X-ray emission from A\,30 can be modeled using 
an optically thin plasma emission model.  
The Astrophysical Plasma Emission Code (APEC) v1.3.1 available 
within XSPEC v12.3.0 \citep{Arnaud96} was used for the spectral 
analysis of the EPIC spectra, adopting the chemical abundances 
of the stellar wind derived from our non-LTE model listed in 
Table~\ref{table:stellarparameters}.

Alternatively we may consider charge transfer reactions between heavy 
ions in the stellar wind and material from the hydrogen-poor knots and 
dust in the central regions of A\,30, as is typically detected in 
comets in the solar system \citep[e.g.,][]{Lisse_etal96,DET97}, as well 
as in a broad variety of astrophysical objects including the interstellar 
medium, stellar winds and galaxies \citep[see][for a review]{Dennerl10}.  
The X-ray emission associated with charge transfer reactions in comets 
can be described by emission lines of the ions involved in these reactions 
with little or negligible continuum (``bremsstrahlung'') emission.  
Our model of charge transfer reactions for the X-ray emission from 
A\,30 will consist of the emission lines in the spectral range 
0.2--0.7 keV of the most important species in the stellar wind of 
the CSPN of A\,30: 
C~{\sc v} 0.31 keV, C~{\sc vi} 0.37 keV, 
N~{\sc vi} 0.43 keV, N~{\sc vii} 0.50 keV, 
O~{\sc vii} 0.57 keV, and O~{\sc viii} 0.65 keV.

\subsubsection{X-ray Absorption Model for A\,30}

The X-ray emission from the hot plasma in A\,30 is certainly absorbed, 
but the nature, properties, and amount of the absorbing material need 
to be elaborated.  
The extinction towards the central star of A\,30 seems to be 
relatively high: \citet{Cohen_etal77}, \citet{Greenstein81}, and 
\citet{Jeffery95} derived interstellar extinctions of $A_V$=0.9$\pm$0.1 
mag, $E_{B-V}$=0.30 mag, and $A_V$=1.18 mag, respectively, whereas we 
have determined it to be $E_{B-V}$=0.18$\pm$0.05 mag.  
As for the extinction towards the central knots, \citet{WLB03} 
derived $c$(H$\beta$)=0.60.  
All these values are in sharp contrast with the low, almost negligible 
extinction affecting the outer shell of A\,30 \citep{GM96}, thus 
suggesting that the origin of the extinction is mostly circumstellar.

\begin{figure*}
\begin{center}
\includegraphics[angle=0,scale=0.4]{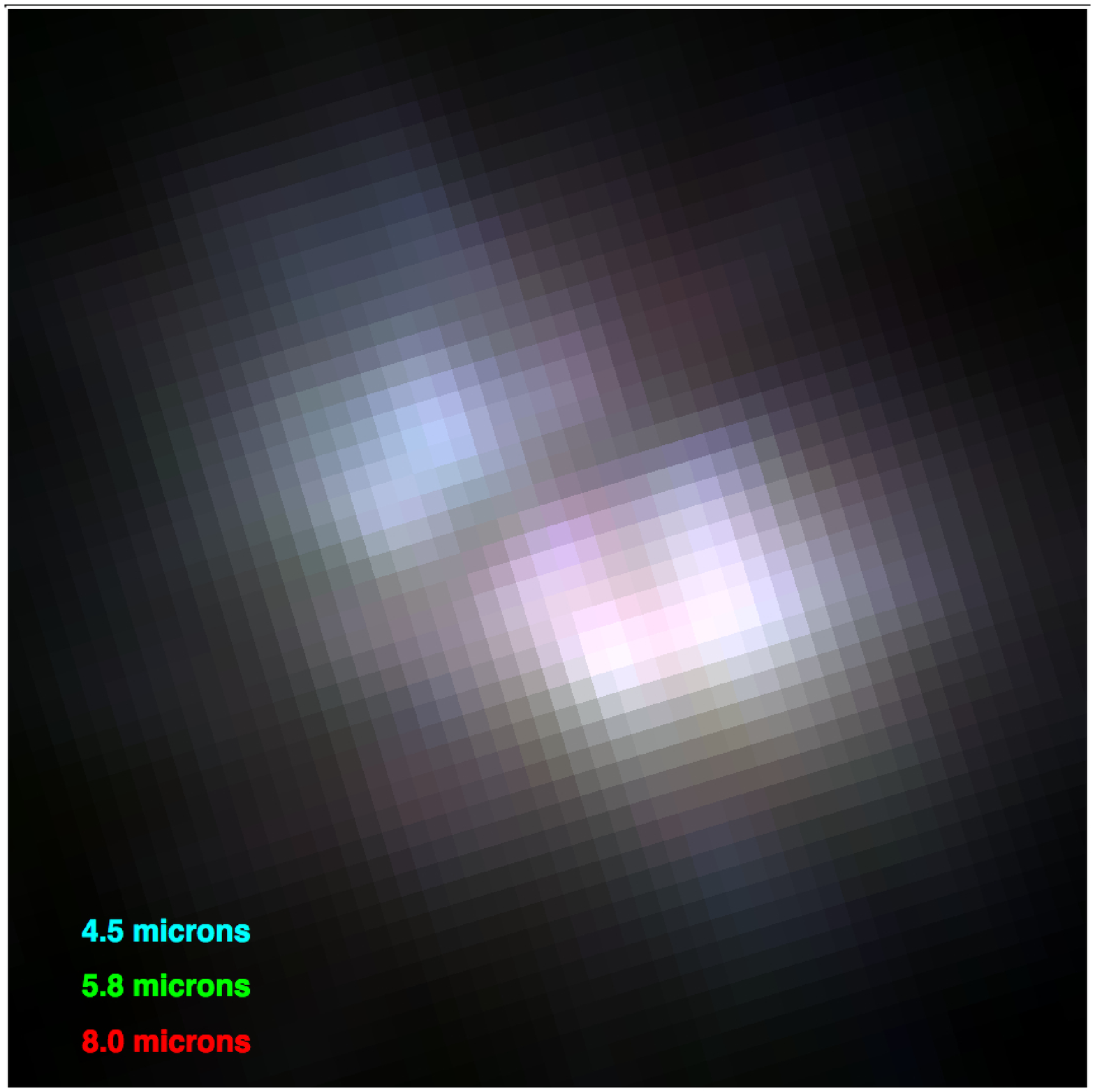}~~
\includegraphics[angle=0,scale=0.4]{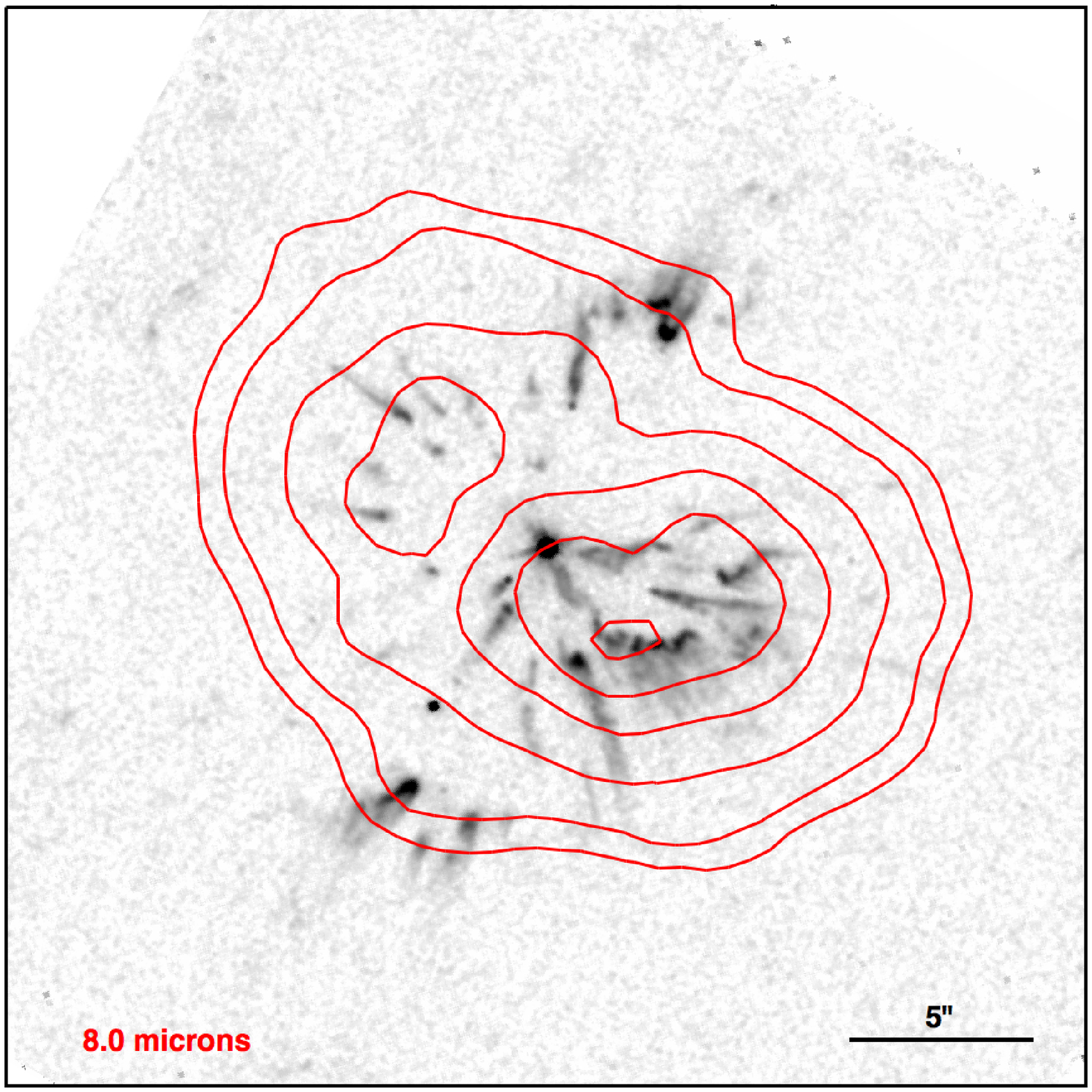}
\end{center}
\caption{
Comparison of \emph{HST} [O~{\sc iii}] and \emph{Spitzer} IRAC 
images of the central regions of A\,30.  
{\it (left)} 
Color-composite \emph{Spitzer} 4.5 $\mu$m (blue), 5.8 $\mu$m (green), 
and 8.0 $\mu$m (red) image of A\,30 obtained using the IRAC observations 
21967616 (PI: G.\ Fazio).  
The subtle variations in the color indicates that the spectral energy 
distribution in the IRAC bands of the dust emission is rather flat, 
with a subtle 4.5 $\mu$m excess in the eastern regions of the equatorial 
ring and some 8.0 $\mu$m excess at the location of the CSPN.  
{\it (right)} 
Grey-scale of the \emph{HST} WFPC2 [O~{\sc iii}] image of A\,30 overplotted 
by \emph{Spitzer} IRAC 8.0 $\mu$m contours.  
North is up, east is left.  
The spatial scale is shown on the \emph{HST} [O~{\sc iii}] image.  
}
\label{Spitzer}
\end{figure*}

The presence of circumstellar dust is indeed revealed by 
mid-IR and near-IR observations of the innermost regions 
of A\,30 \citep[e.g.,][]{Borkowski_etal94,PR-L07}. 
More recent \emph{Spitzer} archival images \citep{Hart_etal11} clearly 
show the spatial coincidence between the mid-IR emission in the IRAC 
bands and the disk and bipolar jet features in \emph{HST} [O~{\sc iii}] 
images (Figure~\ref{Spitzer}). 
The correspondence between the spatial distributions of dust and 
born-again ejecta and the anomalously high carbon composition of 
the dust \citep{Greenstein81,Jeffery95} suggest that this dust is 
formed by material ejected during the VLTP event.  
Consequently we will assume that the absorbing material has a 
chemical composition similar to that of the hydrogen-poor knots, 
i.e., H:He:C:N:O:Ne = 1:11.2:0.47:0.29:1.33:0.56 by number 
\citep{WLB03}.  
Although this absorbing material presents noticeable absorptions at the 
energy of the carbon and oxygen K shells, we must note that at the 
spectral resolution of the EPIC instruments, the overall shape of the 
absorption curve of this metal rich material is similar, within a factor 
1.5, to that of typical interstellar material.

As for the amount of absorbing material, we have performed a simultaneous 
spectral fit of the \emph{XMM-Newton} EPIC-pn and EPIC-MOS, and \emph{ROSAT} 
PSPC spectra of A\,30 using an APEC optically thin plasma emission model 
with stellar wind abundances absorbed by material with the abundances of 
the hydrogen-poor knots.  
The fit is not impressively good ($\chi^2/$d.o.f.=199.6/96$\sim$2.1), 
but it clearly constrains the hydrogen column density at a value 
$\sim$2$\times$10$^{15}$ cm$^{-2}$, with a 3-$\sigma$ upper limit 
$\leq$1$\times$10$^{16}$ cm$^{-2}$, for a plasma temperature of 0.070$\pm$0.005
keV (Table~\ref{table:apec}) and observed flux and intrinsic luminosity in the 
0.2-1.5 keV energy range of 1.0$\times$10$^{-13}$ erg~cm$^{-2}$~s$^{-1}$ and 
4.4$\times$10$^{31}$ erg~s$^{-1}$, respectively.  
Much higher column densities are proscribed by the emission detected in 
the softest energy channels of the EPIC spectra and very notably of the 
\emph{ROSAT} PSPC spectrum.  
We note that the value of the hydrogen column density derived from 
this fit is significantly smaller than those typical of interstellar 
material for PNe, in the range 10$^{19}$--10$^{22}$ cm$^{-2}$, because 
the content in helium, carbon and oxygen assumed for the absorbing 
material is much higher than for the interstellar gas.  
%

Incidentally, we note that the normalization factors of the EPIC and PSPC 
spectra are similar, within 10\%.  
Given the relative calibration uncertainties, the total X-ray fluxes 
from A\,30 determined by \emph{ROSAT} PSPC on 1993 May and by 
\emph{XMM-Newton} EPIC on 2009 October are consistent with each other 
and imply little long-term variability.


\subsubsection{Spectral Fits for Plasma Emission}

The absorbed APEC optically thin plasma emission model provides a reasonable 
fit to the EPIC-pn and EPIC-MOS spectra of the diffuse emission for a best-fit 
value of the temperature $kT$=0.068$^{+0.002}_{-0.005}$ keV 
(=0.79$\times$10$^6$~K) at a fixed hydrogen column density 
of 2$\times$10$^{15}$ cm$^{-2}$ (Table~\ref{table:apec}).   
The observed flux in the 0.2-1.5 keV energy range is 
2.8$\pm0.9\times$10$^{-14}$ erg~cm$^{-2}$~s$^{-1}$, and the 
intrinsic luminosity is $\sim$1.3$\times$10$^{31}$ erg~s$^{-1}$.

This same model does not provide a good fit (reduced $\chi^2\sim 3$) to 
the EPIC-pn and EPIC-MOS spectra of the central source of A\,30 because 
it cannot reproduce the emission peak at $\sim$0.37 keV.  
The addition of an emission line at this energy improves significantly 
the fit (Table~\ref{table:apec}) for a fixed hydrogen column density 
$N_{\rm H}$=2$\times$10$^{15}$ cm$^{-2}$ and best-fit value of 
$kT$=0.068$\pm$0.003 keV (=0.79$\times$10$^6$~K).  
For this model we derive an observed flux in the 0.2-1.5 keV energy 
range of 7.2$^{+3.0}_{-1.8}$$\times$10$^{-14}$ erg~cm$^{-2}$~s$^{-1}$, and 
an intrinsic luminosity $\sim$3.1$\times$10$^{31}$ erg~s$^{-1}$.

The inclusion of an emission line at 0.37 keV is highly indicative 
of increased emission of the C~{\sc vi} line at 33.7 \AA\ which can be 
attributed to an enhancement of the carbon abundances or to a higher 
temperature plasma component.  
Neither possibilities seem to work: the enhancement of carbon 
abundances increases both the emission of the C~{\sc v} and 
C~{\sc vi} lines without a net improvement of the spectral fit, 
whereas the inclusion of a higher temperature component produces 
noticeable emission above 0.45 keV which is not supported by the 
observed spectrum.

\subsubsection{Spectral Fits for Charge Transfer Reactions}

In this case, the spectral model consists of the emission lines of 
C~{\sc v} 0.31 keV, C~{\sc vi} 0.37 keV, N~{\sc vi} 0.43 keV, 
N~{\sc vii} 0.50 keV, O~{\sc vii} 0.57 keV, and O~{\sc viii} 0.65 keV 
at a fixed absorption hydrogen column density of 3$\times$10$^{15}$ 
cm$^{-2}$.  
This model also yields good fits for the emission from the CSPN and 
the diffuse component (Table~\ref{table:cex}).  
The intensities of the different emission lines for the best fit models 
listed in Table~\ref{table:cex} indicate that C~{\sc v} 0.31 keV 
($\equiv$40.2 \AA) is the prevalent line, with a significant contribution 
of the C~{\sc vi} 0.37 keV ($\equiv$33.7 \AA) in the central source.  
Unfortunately, the RGS spectral coverage is limited to the 5--38 \AA\ 
wavelength range and these observations did not provide confirmation 
of the prevalence of the C~{\sc v} line.  
Small contributions of the O~{\sc vii} 0.57 keV ($\equiv$21.8 \AA) 
line to the diffuse emission and central source are also derived 
from these fits.  
On the other hand, the contribution from emission lines of nitrogen is 
found to be negligible both for the diffuse component and for the CSPN 
in agreement with its lower abundance in the stellar wind and born-again 
ejecta.


\section{Discussion}

The new \emph{Chandra} and \emph{XMM-Newton} observations of A\,30 have 
confirmed the extremely soft X-ray emission previously detected by 
\emph{ROSAT} PSPC and HRI \citep{ChuHo95,CCC97} and resolved the X-ray 
emission into a point source and diffuse emission.  
The comparison between the X-ray and optical images of A\,30 allows 
us to unambiguously associate the X-ray point source with its CSPN.  
On small angular scales, as probed by \emph{Chandra} and earlier 
suggested by \emph{ROSAT} HRI images \citep{CCC97}, the spatial 
coincidence of the diffuse X-ray emission with the [O~{\sc iii}] 
bright knots (Figure~\ref{img.xopt}-{\it left}) strongly supports 
the association of the diffuse X-ray emission with the innermost 
hydrogen-deficient knots. 
On larger angular scales, as probed by \emph{XMM-Newton}, the diffuse 
X-ray emission pervades the central regions of the nebula and fills 
the cloverleaf structure (Figure~\ref{img.xopt}-{\it right}).  
No diffuse X-ray emission is found in the gap between the 
cloverleaf structure and the edge of the outer round shell.

We next describe in detail the formation and evolution of A\,30 in order to 
assess which processes can be involved in the production of the diffuse and 
point-source X-ray emission from this nebula.

\subsection{Formation and Evolution of A\,30}

The outer shell of A\,30 formed $\sim$12,500 yr ago from an ordinary AGB 
wind and was shaped by the stellar radiation field and hydrogen-rich fast 
wind of the post-AGB star.  
We see now the relics of this evolution: a large, nearly spherical 
shell of low density and moderate expansion velocity.

About 850 yr ago, the CSPN of A\,30 experienced a VLTP episode which led 
to the sudden ejection of highly processed hydrogen-poor, carbon-rich 
material. 
At that time, the stellar envelope expanded and the star returned to 
the AGB phase, but shortly afterward the envelope contracted and the 
star moved to the post-AGB evolutionary track in time-scales as short 
as 5--20 yr \citep{IM95,Miller_etal06}.  
We can thus expect that, during the AGB phase after the VLTP episode, 
a new carbon-rich wind of low speed blew into the nebular cavity.  
Later, as the star contracted, this carbon-rich wind accelerated 
up to the present terminal velocity of 4000 km~s$^{-1}$ revealed 
by UV spectra of the CSPN of A\,30.

This description of the born-again event and subsequent evolution of 
the stellar wind of A\,30 is limited by our poor understanding of the 
born-again and post-born-again evolution, based on the very small 
sample of known born-again PNe, besides A\,30: 
A\,58 \citep[a.k.a.\ V605\,Aql,][]{Seitter87}, 
A\,78 \citep{JF83}, and 
Sakurai's object \citep[a.k.a.\ V4334\,Sgr,][]{DB96}.  
The duration and properties of the wind during the AGB phase after 
the born-again event are particularly poorly known.
In Sakurai's object, a post-born-again stellar wind with mass-loss rate up 
to 1.6$\times$10$^{-5}$ $M_\odot$~yr$^{-1}$ \citep{Tyne_etal02} and terminal 
velocity $\sim$670 km~s$^{-1}$ \citep{Eyres_etal99} is detected just a few 
years after the born-again event.  
In A\,58, the properties of the present stellar wind, with a diminished 
mass-loss rate of 1$\times$10$^{-7}$ $M_\odot$~yr$^{-1}$ and a terminal 
velocity $\sim$2500 km~s$^{-1}$ \citep{Clayton_etal06}, indicate that 82 
yr after the VLTP event the star has already returned to the post-AGB 
evolutive track.  
It thus seems that, after a born-again event, a post-born-again 
wind with mass-loss rates 10$^{-5}$--10$^{-6}$ $M_\odot$~yr$^{-1}$ 
and terminal velocity of a few hundred km~s$^{-1}$ can be expected 
for a short phase (few years).  
This wind is immediately superseded by a fast stellar wind.


The post-born-again and present fast stellar winds will overtake and 
ablate the hydrogen-poor clumps ejected during the born-again event 
to produce fanning tails and cavities similar to those predicted by 
\citet{SL04}.  
This interpretation is consistent with the nebular features moving at speeds 
as high as 200 km~s$^{-1}$ that have been associated with the fanning tails of 
the hydrogen-poor bipolar knots described by \citet{ML96} and \citet{CCC97}.  
A close inspection of the [O~{\sc iii}] echelle spectra presented by 
\citet{ML96} and those available in the ``SPM Kinematic Catalogue of 
Galactic Planetary Nebulae'' \citep{Lopez_etal12} reveals faint 
features along the central line of sight and associated with some 
cloverleaf features moving at speeds up to 400 km~s$^{-1}$.  
It is worthwhile to note that similar features, moving at speeds 
of 250 km~s$^{-1}$, are found in the hydrogen-poor ejecta of A\,78 
\citep{Meaburn_etal98}.

These features can be interpreted as signatures of the post-born-again 
wind in which the hydrogen-poor knots are embedded.  
We note that the filaments of the cloverleaf structure are not detected 
in the H$\alpha$ image, thus suggesting that they consist of hydrogen-poor 
material.  
The distance of the outermost cloverleaf filaments to the central star 
implies a linear size of 1.1$\times$10$^{18}$ cm that, in conjunction 
with the age of 850 yr derived in Sect.\ 3, result in an expansion 
velocity $\sim$420 km~s$^{-1}$, very similar to the expansion velocity 
of the post-born-again stellar wind of $\sim$400 km~s$^{-1}$.  
This suggests that the post-born-again wind may have had a 
foremost contribution in blowing the cloverleaf structure.

\subsection{Origin of the Diffuse X-ray Emission}



The ISW model of PN formation predicts the production of a ``hot bubble'' 
\citep{KPF78} filled with X-ray-emitting shocked stellar wind as the 
result of the interaction of the CSPN fast stellar wind ($v_\infty > 10^3$ 
km~s$^{-1}$) with the previous slow and dense AGB wind ($v_{\mathrm{AGB}} 
\sim 10$~km~s$^{-1}$).  
X-ray observations of PNe \citep[e.g.,][]{Kastner_etal00,Chu_etal01} 
have detected the diffuse emission from the shocked stellar wind 
inside hot bubbles of PNe with X-ray luminosities in the range
7$\times$10$^{29}$--2$\times$10$^{32}$~erg~s$^{-1}$ (Ruiz et al., in
preparation).  
The luminosity and temperature of the X-ray-emitting gas in PNe are 
satisfactorily reproduced by one-dimensional radiative-hydrodynamic 
models of the formation of PNe which include thermal conduction at 
the interface between the shocked wind and the cold outer shell 
\citep{SSW08}.  
Both observations and models indicate that the X-ray luminosity 
of PNe decays in short time-scales as the CSPN fades and the 
nebula expands.  
For a large, evolved PN such as A\,30, no diffuse X-ray emission from 
the hot bubble enclosed by the AGB wind is expected \citep{GCG00}.  
Furthermore, we expect the hot bubble to collapse toward the star  
as it is not supported any longer by the post-AGB wind once it 
ceased after the VLTP episode.  

The correspondence between the spatial distribution of the X-ray 
emission in A\,30 and the cloverleaf structure suggests that the 
same physical mechanism that generates the X-ray-emitting gas is 
blowing these petal-like features.  
The post-born-again stellar wind, with a terminal velocity of $\sim$400 
km~s$^{-1}$ for A\,30 and a mass-loss rate that could have reached up to 
10$^{-5}$--10$^{-6}$ $M_\odot$~yr$^{-1}$, may provide the power to generate 
the observed X-ray emission.  
We note, however, that hot bubbles of shocked stellar winds 
do not form at these wind speeds because of the very efficient line 
cooling of hydrogen-poor, carbon and oxygen-enriched material 
\citep{ML02,Sandin_etal11}.  

The large momentum and mechanical luminosity of the current fast 
stellar wind of A\,30 can result in strong interactions with the 
ejecta in hydrogen-poor knots and have the potential to power the 
observed diffuse X-ray emission.
The volume of this cavity, 2.2$\times$10$^{54}$ cm$^3$, and the 
emission measure of the extended component, 2.9$\times$10$^{49}$ 
cm$^{-3}$ (Table~\ref{table:apec}), imply a gas
density $N_e=0.006 \times (\epsilon/0.1)^{-1/2}$ cm$^{-3}$, 
%
%
%
%
%
where the gas filling factor $\epsilon$ is presumably low.  
The total mass of the X-ray-emitting gas would be 
2$\times$10$^{-5} \times (\epsilon/0.1)^{1/2}$ $M_\odot$ and, 
%
%
%
%
for a time scale of 850 yr, the averaged mass injection rate of 
X-ray-emitting gas is 3$\times$10$^{-8} (\epsilon/0.1)^{1/2}$ $M_\odot$
yr$^{-1}$, which is consistent with the mass-loss rate of the present 
stellar wind, 2$\times$10$^{-8}$ $M_\odot$~yr$^{-1}$, derived in 
Sect.\ 2.  

The temperature of the X-ray-emitting plasma detected in A30, 
$kT \sim 0.07$~keV, is much too low compared to the post-shock 
temperature expected from a stellar wind with a terminal velocity 
of 4000 km~s$^{-1}$.  
Even when heat conduction is considered in models of hot bubbles 
in PNe, the expected temperature of the X-ray-emitting gas for a 
wind with such a terminal velocity is in the range 0.13--0.43 keV 
\citep{SSW08}.  
We note here that the hydrogen-poor, carbon- and oxygen-rich nature of 
the X-ray-emitting gas implies very efficient line cooling \citep{ML02} 
which can be invoked in conjunction with heat conduction to reduce the 
temperature of the shocked stellar wind \citep{Steffen_etal12}.

Alternatively, the origin of such low temperatures lies in the complex 
interactions between the post-born-again and present fast stellar winds 
of A\,30 and the hydrogen-poor ejecta, as illustrated by the hydrodynamic 
simulations presented by \citet{SL04} and \citet{Pittard07}.  
The hydrogen-poor knots are photoevaporated by the strong UV radiation 
flux of the CSPN and subsequently swept-up by the fast stellar winds,   
which also entrain the H-poor gas ejected after the VLTP episode to 
form the petal-like features seen in the outer shell.  
As a result, material with high metal content can be transferred 
to the shocked stellar wind in three different ways 
\citep[see][]{Arthur07,Pittard07}: 
(1) hydrodynamic ablation, 
(2) conductively-driven thermal evaporation, and 
(3) photoevaporation.  
By increasing the density and damping the velocity of the stellar wind, 
these processes will lower the temperature of the shock-heated stellar 
wind \citep[e.g.,][]{Arthur12}, which will be further reduced by the 
efficient cooling of the high metal abundances of the plasma.



\subsection{Origin of the X-ray Emission at the CSPN of A\,30}

If we concur that the point-source of X-ray emission at the CSPN of A\,30 
originates from a hot plasma, its emission measure, as derived from the 
spectral fits in Sect.\ 4.3.3, is 4.8$\times$10$^{49}$ cm$^{-3}$.  
Since the emitting region is unresolved by \emph{Chandra}, 
the emission volume must be smaller than a sphere with 
radius 0\farcs5, i.e., $\sim$1$\times$10$^{49}$ cm$^3$.  
These figures imply a density of 4$\,\epsilon^{-1/2}$ cm$^{-3}$ and 
a total mass of 6$\times$10$^{-8}\,\epsilon^{1/2}~M_\odot$.  

The exact mechanism responsible for the production of X-ray-emitting hot 
plasma at the CSPN of A\,30 is uncertain.  
Several possibilities are considered in the following.

\subsubsection{Photospheric Emission of the Hot CSPN}

The photospheric emission from hot ($T_{\rm eff}>100,000$~K) CSPNe 
can be detectable in the soft X-ray domain \citep{GCG00}.  
Since the CSPN of A\,30 has an effective temperature of 115,000~K, 
its X-ray emission may be attributed to photospheric emission. 
Figure\,\ref{fig:sed} shows the observed SED together with the stellar 
model presented in Sect.\ 2.  
Whereas there is a good match between the observed SED and predicted 
stellar model in the UV, optical and IR regimes, besides the 2470 \AA\ 
UV bump, the X-ray flux predicted by the model is ten orders of 
magnitude lower than the X-ray flux observed by \emph{XMM-Newton}.  
According to the model, the photospheric X-ray emission is mainly 
blocked by the bound-free and K-shell opacities from C, N and O. 
Hence Fig.\,\ref{fig:sed} leads us to the firm conclusion that the 
stellar photosphere of the CSPN of A\,30 cannot be the origin of 
the observed X-rays.

\subsubsection{Shocks Within the Stellar Wind}

The stellar winds of CSPNe are radiatively driven, i.e., the stellar wind's 
momentum is provided by radiation pressure on spectral lines.  
We can thus expect that the hydrodynamic instability, which is inherent 
to radiatively driven stellar winds, will lead to shocks embedded in the 
stellar wind and produce X-ray emission as for the stellar winds of OB 
and Wolf-Rayet stars \citep[e.g.,][]{Feldmeier_etal97}.  
The X-ray luminosity of the stellar wind of O stars is 
found to scale with the bolometric luminosity as 
$L_{\rm X} \approx 10^{-7}\ L_{\rm bol}$ \citep[][]{Berghofer_etal97}, 
with a scatter of about one order of magnitude.  
For this canonical relationship, the stellar luminosity of the CSPN of A\,30, 
$\log L_{\rm bol}/L_\odot = 3.78$ (Table\,\ref{table:stellarparameters}) 
would imply an X-ray luminosity from the wind-shock emission of 
$10^{-3.2}\ L_\odot$ or 3.5$\times$10$^{30}$\,erg\,s$^{-1}$.  
The expected X-ray emission from wind shocks in the CSPN of A\,30 
is $\sim$50 times lower than observed.

\begin{figure*}[t]
\begin{center}
\includegraphics[scale=0.6]{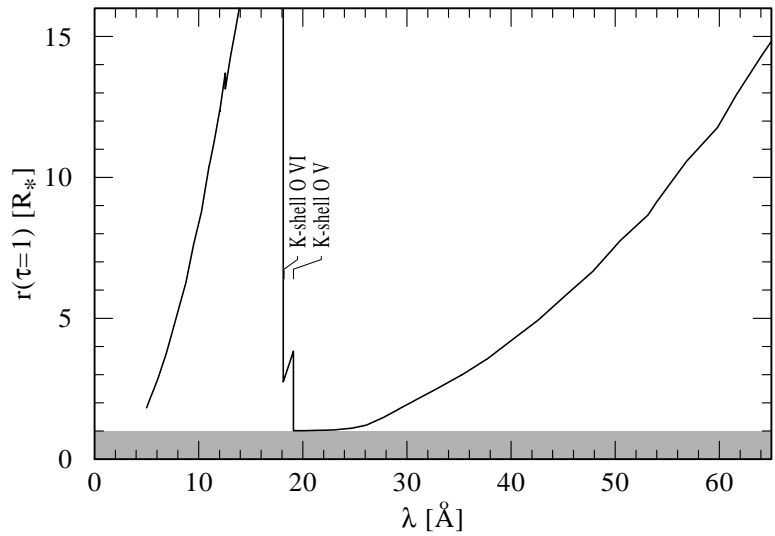}
\end{center}
\caption{
Radius where the radial optical depth reaches unity as a function 
of wavelength according to our PoWR model for the CSPN of A\,30 
(Table\,\ref{table:stellarparameters}).  
The gray band marks optical depth values below unity.  
The location of the O~{\sc v} and O~{\sc vi} K-shell have been labeled.  }
\label{fig:rtau1}
\end{figure*}

Moreover, the aforementioned $L_{\rm X}$-$L_{\rm bol}$ relationship is valid 
for (massive) O stars, but Wolf-Rayet stars are much fainter in the 
X-ray domain than O stars.  
Indeed, the first positive detection of faint X-rays from an WC/WO type 
star has been only recently reported \citep[][]{Oskinova_etal09}.  
The most plausible explanation for the X-ray faintness of Wolf-Rayet 
stars is that their dense winds are very opaque to X-rays, which are 
presumed to be produced by shocks located in the zone of strong wind 
acceleration, deep in the wind at a few stellar radii only.
To assess whether the wind-shock X-rays could emerge from the [WC]-type 
central star A\,30, we plot in Figure~\ref{fig:rtau1} the prediction of 
our PoWR model of the CSPN of A\,30 for the radius where the optical 
depth reaches unity vs.\ wavelength.  
The wind is basically transparent down to the photosphere at 20 \AA\ 
(0.62 keV), but at longer wavelengths, the opacity increases and the 
wind stays optically thick out to 7 stellar radii above 50\,\AA\ 
(0.25 keV).  
The softest X-rays are expected to suffer the strongest attenuation, 
contrary to the properties of the observed X-ray spectrum.  
We thus conclude that the X-ray emission from the unresolved central 
source of A\,30 is not due to shocks embedded in its stellar wind.

\subsubsection{Born-again Hot Bubble}

The mechanical luminosity ($\frac{1}{2}~\dot{M}~v_\infty^2$) of the 
present stellar wind of the CSPN of A\,30 derived from our PoWR model 
(cf.\ Table~\ref{table:stellarparameters}), $\approx 1.0 \times 10^{35}$ 
erg~s$^{-1}$, is $\sim$3000 times larger than the observed X-ray 
luminosity.  
A small fraction of this mechanical luminosity would be able 
to power the X-ray emission which is observed at the central 
source of A\,30.  
As for the diffuse X-ray emission detected in other PNe, the present stellar 
wind can be heated when it rams into previously ejected slower material to 
form a hot bubble.  
The situation in A\,30 is different from other PNe due to its born-again 
history: while the hot bubble of the old, large PN has faded out, a 
``born-again hot bubble'', powered by the present wind, may be responsible 
of the X-ray emission at the central source.  
The observed X-ray temperature is however much lower than the expectations 
for a shocked stellar wind with terminal velocity of 4000 km~s$^{-1}$.  
Heat conduction and/or mass loading most certainly need to be 
invoked to cool the plasma to the observed low temperature. 

Whereas the origin of the X-ray emission of the CSPN of A\,30 in a 
born-again hot bubble is able to explain its luminosity, there are 
critical arguments against this scenario.  
First, the existence of a hot bubble will trap the present fast stellar 
wind and prevent its interaction with the hydrogen-poor knots; however, 
this problem can be mitigated by the post-born-again wind which may have 
contributed mass to the hot plasma.
The second argument against the hot bubble scenario is critical.  
Hot bubbles in PNe fill the whole volume of the innermost shells 
detected in the optical (Ruiz et al., in preparation), but the 
X-ray emission arising from the location of the CSPN of A\,30 is 
unresolved by \emph{Chandra}.  
For a radius of 1.3$\times$10$^{16}$ cm (i.e., 0\farcs5 at the 
distance of A\,30), a hot bubble expanding with the velocity of 
the post-born-again wind (400 km~s$^{-1}$) requires a time of 
just 10.5 yr, implying that the transition from the post-born-again 
wind to the present fast stellar wind occurred some time in 1999.  
This is at odds with the similar X-ray fluxes detected by \emph{ROSAT} PSPC 
on 1993 May and \emph{XMM-Newton} EPIC on 2009 October, and is definitely 
not supported by the quick transition between the post-born-again and fast 
stellar winds observed in A\,58 \citep{Clayton_etal06}.  
%
We conclude that an unresolved born-again hot bubble is 
difficult to sustain as the origin of the X-ray emission 
at the CSPN of A\,30.

\subsection{Charge Transfer Processes}

The spectral analysis described in Sect.\ 4.3.4 and summarized in 
Table~\ref{table:cex} implies that the observed \emph{XMM-Newton} 
EPIC spectra of A\,30 can be reproduced by a model consisting only 
of C~{\sc v}, C~{\sc vi} and O~{\sc vii} lines.  
Such spectral characteristics are expected if the X-ray emission is 
produced by charge transfer from the ions of the stellar wind to 
material of the hydrogen-poor knots or onto the surfaces of dust 
grains, as in solar system comets \citep{Kras97,Lisse_etal99,Dennerl10}.

As for charge transfer reactions between the stellar wind and neutral 
material of the hydrogen-poor knots, these are unlikely as we expect 
the knots to be surrounded by dense ionized outflows 
\citep{HF84,Borkowski_etal95} that will impede the stellar wind to 
penetrate deep into the neutral core of the knot.   
Ion-ion charge transfer processes, considered for the production of X-ray 
emission in the winds of hot stars \citep{Pollock12}, is plausible as the 
large kinetic energy of the ions in the stellar wind of A\,30 can 
overcome the Coulumb repulsion of carbon and oxygen ions in the knots.  
On the other hand, neutralization of highly ionized ions on the surface 
of dust grains \citep[e.g.,][]{BS97} can operate because the production 
of hot dusty plasmas can be expected in the interaction of the fast wind 
with the born-again ejecta.

We note that these processes must compete with others that can be 
pressumably important, such as recombinations of the ions of the 
stellar wind that will reduce the number of available highly-ionized 
species.  
An upper limit of the X-ray luminosity produced by charge transfer 
reactions can be computed from the following relation \citep{Dennerl10}: 
\begin{equation}
L_{\rm X} \sim v_\infty \, n_{\rm w} \, y_{\rm X} \, E_{\rm X} \, S_{\rm k}
\end{equation}
where $v_\infty$ is the wind terminal velocity, 
$n_{\rm w}$ is the wind ion density at the location of the neutral material, 
$y_{\rm X}$ is the fraction of ions capable of releasing an 
X-ray photon in a charge-exchange reaction, 
$E_{\rm X}$ is the photon energy, and 
$S_{\rm k}$ is the cross-section of the hydrogen-poor knots.  
The terminal velocity of the wind ranges from 400 km s$^{-1}$ for the 
post-born-again wind up to 4000 km s$^{-1}$ for the present fast stellar 
wind, whereas their ion density ranges from 10 to 0.001 cm$^{-3}$, 
respectively.  
According to the spectral fit carried out in Sect.\ 4.3.4, the spectrum 
of the diffuse emission of A\,30 is dominated by the C~{\sc v} 0.31 keV 
line. 
Using the energy of the C~{\sc v} line, assuming all carbon 
atoms are found as C~{\sc vi} with a particle fraction of 0.09 
derived from the chemical abundances of the wind, and adopting 
a ring-like structure 5\arcsec\ in radius and 0\farcs5 tall for 
the estimate of the cross-section, the expected X-ray luminosity 
is 2.6$\times$10$^{29}$ erg~s$^{-1}$ for the present fast stellar 
wind and 2.6$\times$10$^{32}$ erg~s$^{-1}$ for the post-born-again 
wind.  
%
%
%
These figures show that the post-born-again wind may produce significant 
X-ray emission through charge transfer reactions, whereas the present fast 
stellar wind can not.

A similar result can be reached for the X-ray emission at the CSPN, where 
the greater density of the stellar wind at close locations of the CSPN 
(0.12 cm$^{-3}$ at 0\farcs5 from the CSPN) is compensated by the smaller 
cross-section, so that the maximum attainable X-ray luminosity would be 
$\sim$5.7$\times$10$^{30}$ erg~s$^{-1}$.  
We conclude that, albeit charge transfer can play a role in the production 
of some the observed X-ray emission, it is unlikely that this mechanism is 
solely responsible for all observed X-ray flux at the CSPN.


\section{Conclusions}

We have used \emph{Chandra} and \emph{XMM-Newton} X-ray observations 
and \emph{HST} multi-epoch archival images of A\,30 to investigate 
the hot gas content and the expansion of the innermost regions of this 
born-again PN.  
Optical and UV high-dispersion spectra have been used, in conjunction 
with the PoWR non-LTE model atmosphere code, to derive the stellar and 
wind properties of its central star.

The large, nearly spherical outer shell of low density of A\,30 meets the 
expectations of the ISW model of PN formation for an old, evolved object, 
in agreement with its kinematical age of 10,000--13,000 yr.  
Since then, the CSPN of A\,30 experienced a VLTP event and ejected 
highly processed material that is detected as large amounts of 
carbon-rich dust and a series of hydrogen-poor, carbon- and oxygen-rich 
knots distributed along an expanding equatorial disk and two bipolar 
outflows.  
The determination of the proper motions of these knots has allowed 
us to derive their expansion rate and to obtain an expansion age of 
850$^{+280}_{-150}$ yr that we interpret as the lapse of time since 
the born-again event.

The CSPN of A\,30 presently exhibits a fast stellar wind with 
terminal velocity $\sim$4000 km~s$^{-1}$ and a low mass-loss rate, 
$\sim$2$\times$10$^{-8} M_\odot$~yr$^{-1}$.  
A careful examination of archival [O~{\sc iii}] echelle data of A\,30 
reveals the occurrence of faint nebular features along the line of sight 
of the CSPN and the cloverleaf-shaped filaments that expand at speeds up 
to 400 km~s$^{-1}$.  
These features can be interpreted as the signatures of the 
post-born-again wind.  
A comparison with the other known born-again CSPNe (A\,58, A\,78 and 
VV4334\,Sgr) suggests that the onset of the post-born-again wind occurred 
soon after the born-again event, with time-scales as short as a few 
years.  
This post-born-again wind is soon superseded by the present fast stellar 
wind.  
Whereas this is an incomplete picture of the evolution of the stellar wind 
during the born-again and post-born-again phases, we note that the present 
model calculations do not offer a more detailed view of this transition.

The above descriptions indicate that A \,30 is a complex object composed 
of a system of three nested winds: 
a post-AGB wind that formed a typical PN shell, 
a medium-speed born-again and post-born-again wind consisting of processed, 
hydrogen-poor material, and 
a present high-speed, hydrogen-poor wind.  
A\,30 is thus a unique system to study the effects of various types of 
wind-interactions.

The exquisite spatial resolution of \emph{Chandra} and unprecedented 
sensitivity of \emph{XMM-Newton} have allowed us to resolve the X-ray 
emission from A\,30 into a point-source at its central star and 
diffuse emission associated with the innermost hydrogen-poor knots and 
with the cloverleaf structure inside the nebular shell.  
The diffuse X-ray emission from A\,30 originates in the interactions 
of the present fast stellar wind and post-born-again wind with the 
hydrogen-poor ejecta.  
After the born-again event, the hydrogen-poor, carbon-rich post-born-again 
wind blew a cavity into the nebula that resulted in the cloverleaf structure.  
The interactions of this wind and the present fast stellar wind with 
clumps of low speed and carbon-rich dust from the born-again event 
result in processes of shock-heating and mass-loading of the stellar 
winds and ablation of the hydrogen-poor knots that produce 
X-ray-emitting plasma.  
Diffuse X-ray emission may also result from charge transfer reactions between 
the stellar winds and the hydrogen-poor ejecta in knots and dust.

The origin of the point-source of X-ray emission at the central star 
of A\,30 is unclear.  
It is unlikely to result from shocks in the stellar wind, as in OB stars, 
or from the hot CSPN photospheric emission.  
The development of a ``born-again hot bubble'' may explain this emission, 
but its small size is puzzling.


\acknowledgments

M.A.G., N.R., and J.T.\ are partially funded by grants AYA\,2008-01934 
and AYA2011-29754-C03-02 of the Spanish MICINN (Ministerio de Ciencia 
e Innovaci\'on) and MEC (Ministerio de Econom\'\i a y Competitividad).  
Y.-H.C.\ and R.A.G.\ acknowledge the support of NASA grants NNX09AU32G 
and SAO~GO01-12029X.  
Funding for this research has been provided by DLR grant 50\,OR\,1101
(LMO).  
We are grateful to an anonymous referee for pointing out the importance 
of charge transfer on dust and for detailed comments on charge transfer 
which significantly improved the manuscript. 
We are very grateful to Ralf Napiwotzki for providing us with the 
optical high-dispersion spectra of the central star of A\,30 and 
to Sarah J.\ Arthur for helpful discussion on charge transfer to 
dust and ions.  



{\it Facilities:} 
\facility{Chandra (ACIS-S)}, 
\facility{HST (WFPC2)}, 
\facility{KPNO}, 
\facility{Spitzer (IRAC)}.
\facility{VLT UT2 (UVES)}, 
\facility{XMM-Newton (EPIC-pn, EPIC-MOS, RGS)}.

\clearpage



\clearpage

\clearpage

\clearpage

\clearpage

\clearpage

\clearpage


\clearpage

\clearpage

\clearpage

\clearpage

\clearpage


\begin{deluxetable}{lrl}
\tablecolumns{3}
\tablewidth{0pc}
\tablecaption{Parameters of the central star of A\,30}
\tablehead{
\multicolumn{1}{l}{Parameter} & 
\multicolumn{1}{r}{Value}     & 
\multicolumn{1}{l}{Comments}
}
\startdata
\multicolumn{3}{l}{Stellar parameters} \\
\hline
$\log\,(L/L_\odot)$   &  3.78      & Adopted \\ 
Clumping factor, $D$  &  10  & Adopted \\
$d$\hspace*{0.88cm}(kpc) &  1.76 & $d \propto L^{1/2}$ \\ 
$E_{B-V}$ (mag)           & 0.18 & \\
$R_\ast$\tablenotemark{a}\hspace*{0.38cm}($R_\odot$) &  0.20 & $R_\ast \propto L^{1/2}$ \\ 
$T_\ast$\tablenotemark{b}\hspace*{0.45cm}(K)       &  115,000 & \\ 
$v_\infty$\hspace*{0.60cm}(km~s$^{-1}$)             & 4000 & \\
$\dot{M}$\hspace*{0.68cm}($M_\odot\,{\rm yr}^{-1}$) & ~~2.0$\times$10$^{-8}$ & $\dot{M} \propto D^{-1/2}L^{3/4}$ \\
\hline
\multicolumn{3}{l}{Abundances (mass fractions)}   \\
\hline
He                             & 0.63   & \\ 
C                              & 0.20   & \\ 
N                              & 0.015  & \\ 
O                              & 0.15   & \\ 
Fe (+ iron group)              & 0.0016 & \\ 
\enddata
\tablenotetext{a}{
~The stellar radius $R_\ast$ refers by definition to the point where the 
radial Rosseland optical depth is 20. }
\tablenotetext{b}{
~$T_\ast$ is defined as the effective temperature related to the radius $R_\ast$.}
\label{table:stellarparameters}
\end{deluxetable}


\begin{deluxetable}{lccccc}
\tablecolumns{6}
\tablewidth{0pc}
\tablecaption{Background-subtracted count rates and net count numbers}
\tablehead{
\multicolumn{1}{l}{Instrument}        & 
\multicolumn{4}{c}{\underline{~~~~~~~~~~~~~~~~~~~Background-subtracted count rate~~~~~~~~~~~~~~~~~~~}}        & 
\multicolumn{1}{c}{\underline{~~Net counts~~}}        \\
\multicolumn{1}{c}{} & 
\multicolumn{1}{c}{190-600 eV$^1$} & 
\multicolumn{1}{c}{190-275 eV} & 
\multicolumn{1}{c}{275-450 eV} & 
\multicolumn{1}{c}{450-600 eV} & 
\multicolumn{1}{c}{190-600 eV\tablenotemark{a}} \\
\multicolumn{1}{c}{} & 
\multicolumn{1}{c}{(counts~ks$^{-1}$)} & 
\multicolumn{1}{c}{(counts~ks$^{-1}$)} & 
\multicolumn{1}{c}{(counts~ks$^{-1}$)} & 
\multicolumn{1}{c}{(counts~ks$^{-1}$)} & 
\multicolumn{1}{c}{(counts)} 
}
\startdata
\multicolumn{6}{c}{Point source and diffuse emission} \\
\hline
EPIC-pn   & 50.4$\pm$1.6~ & 19.7$\pm$0.9 & 28.3$\pm$1.1 & 1.57$\pm$0.31 & 1150$\pm$40 \\
EPIC-MOS1 & ~7.7$\pm$0.6~ & $\dots$      & $\dots$      & $\dots$       &  219$\pm$16 \\
EPIC-MOS2 & ~6.6$\pm$0.5~ & $\dots$      & $\dots$      & $\dots$       &  186$\pm$15 \\
ACIS-S    & 1.80$\pm$0.14 & $\dots$      & $\dots$      & $\dots$       &  172$\pm$13 \\
\hline
\multicolumn{6}{c}{Point source} \\
\hline
EPIC-pn   & 29.7$\pm$1.1~ & $\dots$      & $\dots$      & $\dots$       & 675$\pm$26 \\
EPIC-MOS1 &  4.4$\pm$0.4~ & $\dots$      & $\dots$      & $\dots$       & 140$\pm$12 \\
EPIC-MOS2 &  3.4$\pm$0.3~ & $\dots$      & $\dots$      & $\dots$       & 105$\pm$10 \\
\hline
\multicolumn{6}{c}{Diffuse emission} \\
\hline
EPIC-pn   & 14.3$\pm$0.9~ & $\dots$      & $\dots$      & $\dots$       & 325$\pm$20 \\
EPIC-MOS1 &  2.2$\pm$0.3~ & $\dots$      & $\dots$      & $\dots$       &  71$\pm$9 \\
EPIC-MOS2 &  1.7$\pm$0.3~ & $\dots$      & $\dots$      & $\dots$       & 55$\pm$8 \\
\enddata
\tablenotetext{a}{
~For ACIS-S, the low energy cutoff is not 190 eV, but 200 eV.  }
\label{table:counts}
\end{deluxetable}


\begin{deluxetable}{lccccl}
\tablecolumns{6}
\tablewidth{0pc}
\tablecaption{Best fit parameters for plasma emission models} 
\tablehead{
\multicolumn{1}{l}{Region}          & 
\multicolumn{1}{c}{$N_{\rm H}$}      &
\multicolumn{1}{c}{$kT$}            & 
\multicolumn{1}{c}{EM\tablenotemark{a}} &
\multicolumn{1}{c}{$I_{\rm C VI}$}   & 
\multicolumn{1}{c}{$\chi^2$/d.o.f.} \\
\multicolumn{1}{c}{}                & 
\multicolumn{1}{c}{(cm$^{-2}$)}      & 
\multicolumn{1}{c}{(keV)}           & 
\multicolumn{1}{c}{(cm$^{-3}$)}        & 
\multicolumn{1}{c}{(photon~cm$^{-2}$~s$^{-1}$)} & 
\multicolumn{1}{c}{}                
}
\startdata
A\,30       & (2$\pm$2)$\times$10$^{15}$ &         0.070$\pm$0.005  & 9.2$\times$10$^{49}$  &                       $\dots$ & 2.08 (=199.6/96) \\  
CSPN\tablenotemark{b}    &         2$\times$10$^{15}$ &         0.068$\pm$0.003  & 4.8$\times$10$^{49}$  & (5.0$\pm$0.6)$\times$10$^{-5}$ & 1.11 (=46.5/42) \\   
Diffuse\tablenotemark{b} &         2$\times$10$^{15}$ & 0.068$^{+0.002}_{-0.005}$  & 2.9$\times$10$^{49}$ &                        $\dots$ & 1.38 (=42.7/31  \\ 
\enddata
\tablenotetext{a}{
Emission measure, 
EM = $\int n_{\rm e} n_{\rm ion} dV = 10^{14} 4 \pi d^2 K_{\rm apec}$.
}
\tablenotetext{b}{
Adopted value of the hydrogen column density.
}
\label{table:apec}
\end{deluxetable}


\begin{deluxetable}{lcccccccc}
\tablecolumns{9}
\tablewidth{0pc}
\tablecaption{Best fit parameters for charge-exchange reaction models} 
\tablehead{
\multicolumn{1}{l}{Region}     & 
\multicolumn{1}{c}{$N_{\rm H}$} &
\multicolumn{1}{c}{$I_{\rm C V}$} & 
\multicolumn{1}{c}{$I_{\rm C VI}$} & 
\multicolumn{1}{c}{$I_{\rm N VI}$} & 
\multicolumn{1}{c}{$I_{\rm N VII}$} & 
\multicolumn{1}{c}{$I_{\rm O VII}$} & 
\multicolumn{1}{c}{$I_{\rm O VIII}$} & 
\multicolumn{1}{c}{$\chi^2$/d.o.f.}    \\
\multicolumn{2}{c}{}        & 
\multicolumn{6}{c}{\line(1,0){310}~~~} & 
\multicolumn{1}{c}{}        \\
\multicolumn{1}{c}{}        & 
\multicolumn{1}{c}{(cm$^{-2}$)}        & 
\multicolumn{6}{c}{(photon~cm$^{-2}$~s$^{-1}$)} & 
\multicolumn{1}{c}{}        
}
\startdata
CSPN\tablenotemark{a}    & 2$\times$10$^{15}$ & 7.2$\times$10$^{-5}$ & 6.7$\times$10$^{-5}$ & $<$1$\times$10$^{-8}$ & $<$5$\times$10$^{-9}$ & 1.6$\times$10$^{-6}$ &  4.7$\times$10$^{-8}$ & 1.35 (=37.8/28) \\
Diffuse\tablenotemark{a} & 2$\times$10$^{15}$ & 4.0$\times$10$^{-5}$ & 9.5$\times$10$^{-6}$ & $<$1$\times$10$^{-8}$ & $<$1$\times$10$^{-8}$ & 7.5$\times$10$^{-7}$ & $<$2$\times$10$^{-7}$ & 1.23 (=49.0/40) \\
\enddata
\tablenotetext{a}{
~Adopted value of the hydrogen column density. }
\label{table:cex}
\end{deluxetable}

\end{document}